\documentclass[12pts,fleqn,usenatbib,useAMS]{mnras}

\usepackage{amsmath}
\usepackage{amssymb}
\usepackage{caption}
\usepackage{color}
\usepackage{subfigure}
\usepackage{float}
\usepackage{natbib}
\usepackage[T1]{fontenc}

\RequirePackage{fixltx2e}

\def\la{\mathrel{\hbox{\rlap{\hbox{\lower4pt\hbox{$\sim$}}}\hbox{$<$}}}}
\def\gt{\mathrel{\hbox{$>$}}}

\def\ga{\mathrel{\hbox{\rlap{\hbox{\lower4pt\hbox{$\sim$}}}\hbox{$>$}}}}

\usepackage{journal_names}
\usepackage{color}

\usepackage{url}

\newcommand\farcss{\mbox{$.\!\!\!^{\prime\prime}$}}
\def\farcm{\mbox{.\kern -0.5ex\raisebox{.6ex}{\scriptsize$\prime$}}}

\def\farcss{
 \mbox{ 
  \kern  0.13ex. 
   \kern -0.95ex\raisebox{.6ex}{\scriptsize$\prime\prime$}
  \kern -0.1ex
 }
}

\usepackage{graphicx,times}

\usepackage{float}

\title[Role of stellar encounters in planet formation]{Planet population synthesis: The role of stellar encounters}

\author[Ndugu et al.]{
N.Ndugu,$^{1,2}$\thanks{E-mail: ndugu.nelson89@gmail.com}
O.P. Abedigamba,$^{1,3}$\thanks{E-mail: oyigamba@gmail.com}
G. Andama$^{4}$\thanks{E-mail: gandama@must.ac.ug}
\\
$^{1}$Centre for Space Research, North-West University, \textsc{South Africa}\\
$^{2}$Department of Physics, Muni University, Arua, \textsc{Uganda}\\
$^{3}$Department of Physics, Kyambogo University, Kampala, \textsc{Uganda}\\
$^{4}$Department of Physics, Mbarara University of Science and Technology, 
Mbarara, \textsc{Uganda}\\
}

\pubyear{2022}

\begin{document}
\label{firstpage}
\pagerange{\pageref{firstpage}--\pageref{lastpage}}
\maketitle
\numberwithin{equation}{section}
\numberwithin{figure}{section}
\begin{abstract}
Depending on the stellar densities, protoplanetary discs in stellar clusters undergo: background heating; disc truncation—driven by stellar encounter; and photo-evaporation. Disc truncation leads to reduced characteristic sizes and disc masses that eventually halts gas giant planet formation. We investigate how disc truncation impacts planet formation via pebble-based core accretion paradigm, where pebble sizes were derived from the full grain-size distribution within the disc lifetimes. We make the best-case assumption of one embryo and one stellar encounter per disc. Using planet population syntheses techniques, we find that disc truncation shifts the disc mass distributions to the lower margins. This consequently lowered the gas giant occurrence rates. Despite the reduced gas giant formation rates in clustered discs, the encounter models mostly show as in the isolated field; the cold Jupiters are more frequent than the hot Jupiters, consistent with observation. Moreover, the ratio of hot to cold Jupiters depend
on the periastron distribution of the perturbers with linear distribution in periastron ratio showing enhanced hot to cold Jupiters ratio in comparison to the remaining models. Our results are valid in the best-case scenario corresponding to our assumptions of: only one disc encounter with a perturber, ambient background heating and less rampant photo-evaporation. It is not known exactly of how much gas giant planet formation would be
affected should disc encounter, background heating and photo-evaporation act in a concert. Thus, our study will hopefully serve as motivation for quantitative investigations of the detailed impact of stellar cluster environments on planet formations.

\end{abstract}
%

\section{Introduction}

Despite most stars are born in stellar clusters \citep{2003ARA&A..41...57L} and young stars frequently having circumstellar discs \citep{2000AJ....120.3162L}, there are very few exoplanet collections that are observed in stellar clusters \citep{2019MNRAS.489.4311C}. The paucity of exoplanets in stellar clusters could be due to: (i) photo-evaporation that shortens the lifetime of the embedded discs \citep{1998ApJ...499..758J,2006MNRAS.369..229A,2016MNRAS.457.3593F,2019MNRAS.483.3448M,2021arXiv210505908M}, (ii) stellar encounters which limit the sizes and masses of discs through truncation \citep{2012A&A...538A..10S,2014A&A...565A.130B,2015A&A...577A.115V}, (iii) background heating of protoplanetary discs by the surrounding massive stars \citep{2005ASPC..341..107H,2006ApJ...641L..45K,2013MNRAS.431...63T,2018MNRAS.474..886N} and (iv) few ground-based transit surveys for planets in stellar clusters \citep{2011ApJ...729...63V}. Planets that form in stellar cluster environment encounter the following phases depending on the stellar cluster evolution \citep{2019MNRAS.489.4311C}:

\begin{itemize}
 \item [--] Gas-rich phase: In this early less diverse phase, planet formation takes place in gas rich environment. The high stellar cluster density at this stage means enhanced photo-evaporation/stellar encounters that limits the disc lifetimes/truncates the disc. The proto-planets in this phase are protected from perturbation by the gaseous rich environment.
\item [--] Gas-poor phase: In this phase, due to lack of protection by the damping of the gaseous discs, perturbation, i.e, stellar encounters, planet-planet scattering and tidal interactions are very common.
\item [--] Isolated planetary system phase: In this phase, the emergent planets survived the stellar encounters in the cluster environments depending on the stellar cluster density \citep{2017MNRAS.470.4337C,2019A&A...624A.120V,2019MNRAS.489.2280F}. However, there still exists planet-planet scattering and tidal interaction with planets harboured in high stellar cluster density environments having high mean eccentricities and inclinations \citep{2019MNRAS.489.4311C}.
\end{itemize}

If indeed stars (and planets) form in stellar clusters \citep{2003ARA&A..41...57L,2020MNRAS.495L..86L}, then the current exoplanets are the planets that survived the background heating, photo-evaporation processes and stellar encounter of their birth environments. Therefore, accurate explanation of the current statistics of the exoplanets will require large computational efforts that incorporate all the complex processes within the stellar clusters. Unfortunately, such intensive computational effort in planet formations are still under development \citep{1997ASPC..123..177H,2019BAAS...51c.129L,2021A&A...656A..69E,2021A&A...656A..70E,2021A&A...656A..71S,2021A&A...656A..72B,2021A&A...656A..73S}. As hinted before, for planet forming in stellar clusters, the cluster density is key. At low stellar densities, planet formation is mostly unaffected in open clusters and only the wider planetary systems will be disrupted during the cluster’s life time \citep{2001MNRAS.322..859B} compared to global stellar clusters that have high stellar densities. Photo-evaporation and stellar encounters effects for clusters with varnishing densities therefore have minimal impacts on gas giant planet formation processes within the inner orbits of the clusters \citep[e.g,][]{2000A&A...362..968A,2005MNRAS.364..961T}.

Disc truncation in stellar clusters happens when flybys occur in star forming region with high initial stellar densities, for example in the Orion Nebula Cluster (ONC), NGC 6611 and NGC 2264 \citep{2014A&A...565A.130B,2020RSOS....701271P,2022MNRAS.510.1136P}. Observation of star formation regions show that the properties of discs—particularly their radii, tend to be smaller—in the most dense star-forming regions compared to lower density regions \citep{2012A&A...546L...1D,2017AJ....153..240A,2018ApJ...860...77E,2018MNRAS.478.2700W}. It is however not known if these observed disc-properties are due to disc truncation or photo-evaporation. Stellar flybys have been identified, for example in the recent near-infrared and submillimetre observations (e.g. RW Aur, AS 205, HV Tau and DO Tau, FU Ori, V2775 Ori, and Z CMa) \citep{2020MNRAS.491..504C}. Stellar encounters are even more prominent for the galactic bulge stars, where 80~\% of the stars experienced encounters \citep{2020MNRAS.495.2105M}. Dedicated studies that formulate the impact of stellar flybys on disc properties, exists \citep[e.g,][]{2005A&A...437..967P,2007A&A...462..193P,2012A&A...538A..10S,2014A&A...565A.130B,2019ComAC...6....3B}. For example, \cite{2014A&A...565A.130B} explicitly modelled for parameter space equivalent to the ONC, post encounter disc size depedence on the mass ratio and periastron ratio for a parabolic encounter.

Previous gas giant planet formation studies \citep[e.g,][]{2000A&A...362..968A,2009A&A...505..873F,2013MNRAS.431...63T,2017AJ....154..272H,2018MNRAS.474..886N} hint how stellar clusters suppresses gas giant planet formations. For example, \cite{2000A&A...362..968A} showed that intense photo-evaporation originating from stellar clusters disperses the disc early that eventually reduces the probability of forming gas giant planets. \cite{2009A&A...505..873F} found that stellar fly-by significantly makes alteration to the masses and the orbital parameters of gas giant planet forming within protostellar discs. \cite{2013MNRAS.431...63T,2018MNRAS.474..886N} showed that the background heating from stellar clusters makes the outskirts of protoplanetary discs hot that eventually hinders gas giant planet formation. \cite{2017AJ....154..272H} showed in global clusters, hot Jupiter planets are driven by highly eccentric migration that might eventually deter the accretional efficiency \citep{2010A&A...523A..30B,2010AJ....139.1297L}. With the exception of \cite{2018MNRAS.474..886N} that performed planet population synthesis in cluster environments, most of these models \citep[e.g,][]{2000A&A...362..968A,2009A&A...505..873F,2013MNRAS.431...63T,2017AJ....154..272H} addressed the impact of stellar clusters on planet formation by focusing on specific aspect of planet formation. \cite{2018MNRAS.474..886N} expanded the solar like disc evolution model of \cite{2015A&A...575A..28B} to mimic the disc evolution model for stellar clusters by setting different background temperatures. In the disc, planetary seeds that have reached the pebble transition mass were implanted, indicating that the seeds accrete in the efficient Hill accretion regime \citep{2014A&A...572A.107L}. The planetary embryo accretes pebbles until it reaches the pebble isolation mass, where the pebble flux is shut-off due to interactions of the planet with the disc blocking the flux of pebbles after several 100 ky \citep{2014A&A...572A..35L}, it contracts a gaseous envelope \citep{2014ApJ...786...21P} and then undergoes run-away gas accretion, when the planetary envelope becomes more massive than the planetary core, where the gas accretion rates derived by \cite{2010MNRAS.405.1227M} was used. In addition, planetary migration in the disc was model with type I migration rates following \cite{2011MNRAS.410..293P} prescription and type II migration rate following \cite{2014prpl.conf..667B}. In this paper, we performed a follow up study on our previous work \citep{2018MNRAS.474..886N}. Now instead of focusing on the effect of thermal heating of stellar cluster, we investigate the impact of stellar encounters on planet population by modeling for a fixed background heating, the impact of stellar truncation on disc size \citep{2014A&A...565A.130B} and consequently the disc masses that were initially derived for viscously evolving discs \citep{1974MNRAS.168..603L}.

We also switched from accreting the dominant pebble size to accreting realistically full pebble size distributions in discs onto planetary cores via multi-species pebble accretion paradigm of \cite{2022MNRAS.510.1298A} that built from the existing pebble accretion formalism \citep{2010A&A...520A..43O,2010MNRAS.404..475J,2014A&A...572A.107L}. The gas accretion scheme follows \cite{2021MNRAS.501.2017N}, which show that additional accretion of gas from the horseshoe region allows reduced migration of massive planets compared to the classical gas accretion formalism. In comparison to the previous planet population synthesis study in \cite{2018MNRAS.474..886N,2019MNRAS.488.3625N} that fixed for simplicity reasons some of the important disc properties, we now add more detailed layers of parameters to the disc properties. For example, host star mass, characteristic radius and disc masses are realistically constrained.  As in \cite{2017ASSL..445..339B,2018MNRAS.474..886N,2019MNRAS.488.3625N,2021MNRAS.501.2017N}, the emergent planet population data are compared to the occurrence rates of the exoplanet \citep{2010PASP..122..905J,2011arXiv1109.2497M,2016ApJ...833..145S}.
This work is structured as follows. We discuss: the disc structure (and it's evolution), stellar cluster model, the planet formation model in section~\ref{Methods} and the initial conditions for the planet population synthesis in section~\ref{Initcond}. We discuss the role of stellar cluster encounters on disc properties and planet formation in section~\ref{discmass} and section~\ref{popsynth}, respectively. All the results will be presented showing host disc-size distributions and the final evolutionary stage of planets in terms of planetary mass-orbital distance diagrams, showing the final location and mass as a function of metallicity. In section~\ref{comparison}, we compare our planet formation models to the observed exoplanets from radial velocity (RV) and microlensing surveys. Section~\ref{discuss}  relates the findings to the existing studies and Section~\ref{Conclusion} summarizes the results, putting them into perspectives for future investigations.
\section{Methods}\label{Methods}

\subsection{Protoplanetary Disc Model}
As in \cite{2021A&A...647A..15D}, we set the gas surface density as a function of the radial distance $r$ to the self-similar solution of the viscous evolution equations \citep{1974MNRAS.168..603L}: 
\begin{equation}
    \Sigma_{\rm{gas}}(t =0) = \frac{M_{\rm{disc}}}{2 \pi r_{\rm c}^2} \left(\frac{r}{r_{\rm c}}\right)^{-1}\exp{\left(-\frac{r}{r_{\rm c}}\right)},
\end{equation}
where $M_{\rm{disc}}$ and $r_c$ are the initial disc mass and the characteristic radius, respectively. The dust surface density is then calculated as
\begin{equation}
    \Sigma_{\rm{dust}}(t =0) = Z\cdot\Sigma_{\rm{gas}}(t =0),
\end{equation}
where $Z$ is the global dust-to-gas ratio. In all the models presented in this paper, we consider discs with varying initial mass budget, $M_{\rm{disc}}$ and varying dust-to-gas ratio that are succinctly described in subsection~\ref{Initcond}. We model the disc's characteristic radius, $r_{\rm c}$ as:

\begin{equation}
 r_{\rm c} (t =0) = r^{'}\left({\frac{M_{\rm h,\textasteriskcentered}}{M_{\bigodot}}}\right)^{0.5},
\end{equation} where $r^{'}$ is a constant. The youngest circumstellar discs observed to date have diameters that range from 30~au \citep[e.g.][]{2018ApJ...863...94L} to 120~–~180~au \citep[e.g.][]{2013A&A...560A.103M,2018A&A...615A..83V}. Based on this we choose $r^{'} = 100$~au. For a solar mass host star, the characteristic radius is therefore 100~au. We utilized gas disc model that scales with $\frac{r}{r_{\rm c}}$.  As in \cite{2011ARA&A..49...67W},  we assume that at the characteristic radius of protoplanetary disc,  the surface density of the disc is delineated where it begins to steepen significantly from a power law characteristic radius. It is at the $r_{\rm c}$ where we limit  the outskirt of our disc model. In principle, massive discs are mostly unstable and might fragment. We remind the reader throughout this work that the disc masses used for constructing the disc models are not massive and are therefore gravitationally stable, with the Toomre $Q>1$.

The gas disc evolves viscously following the numerical routines of \cite{2012A&A...539A.148B}, which means that the characteristic radius increases, while the surface density drops with time. The speed of viscous evolution scales with the $\alpha$ parameter that represents viscosity \citep{Shakura1973}. Since the main objective of this paper is to find the global influence of disc truncation on planet formation, and since we adopted the disc truncation fits from \cite{2014A&A...565A.130B} that works for relatively low $\alpha$ cases, we use throughout this paper, a moderate viscosity values; $\alpha=10^{-3}$ and $\alpha=6\times 10^{-4}$ (see the detailed explanation in subsection~\ref{clustermodel}). In addition, the DSHARP survey found $\alpha$-viscosity parameter within the range of $10^{-6}~{\rm to}~10^{-2}$. Our choice of $\alpha$ is thus within the observed limit.

The solid accretion and the migration rates relates to the disc temperature profile. In this work, we model the disc temperature profile following the simple prescription proposed by \citet[][their equation 7 and 8]{2016A&A...591A..72I}. The disc temperature profile described in \cite{2016A&A...591A..72I} feature both viscous and irradiation heating.  Most of the disc is heated by stellar irradiation, leading to a shallow temperature profile $T_{\rm{irr}}\propto r^{-3/7}$. In the inner part of the disc, viscous heating may change the temperature profile to a steeper function of the radial distance, $T_{\rm{vis}}\propto r^{-9/10}$.  We used the stellar luminosity data from \cite{1998A&A...337..403B} to set the irradiation profile of our disc model. The stellar luminosity for the disc model similarly scales as 
\begin{equation}
\frac{L}{L_{\odot}} = 1.05745 \exp\left(-\frac{t}{1.966~My}\right)+0.454424.
\end{equation} 
We note that the disc temperature model used in this study is an analytical model that features regions dominated by viscous and stellar heating, but does not include any opacity transitions.

\begin{figure}
\includegraphics[width=\linewidth]{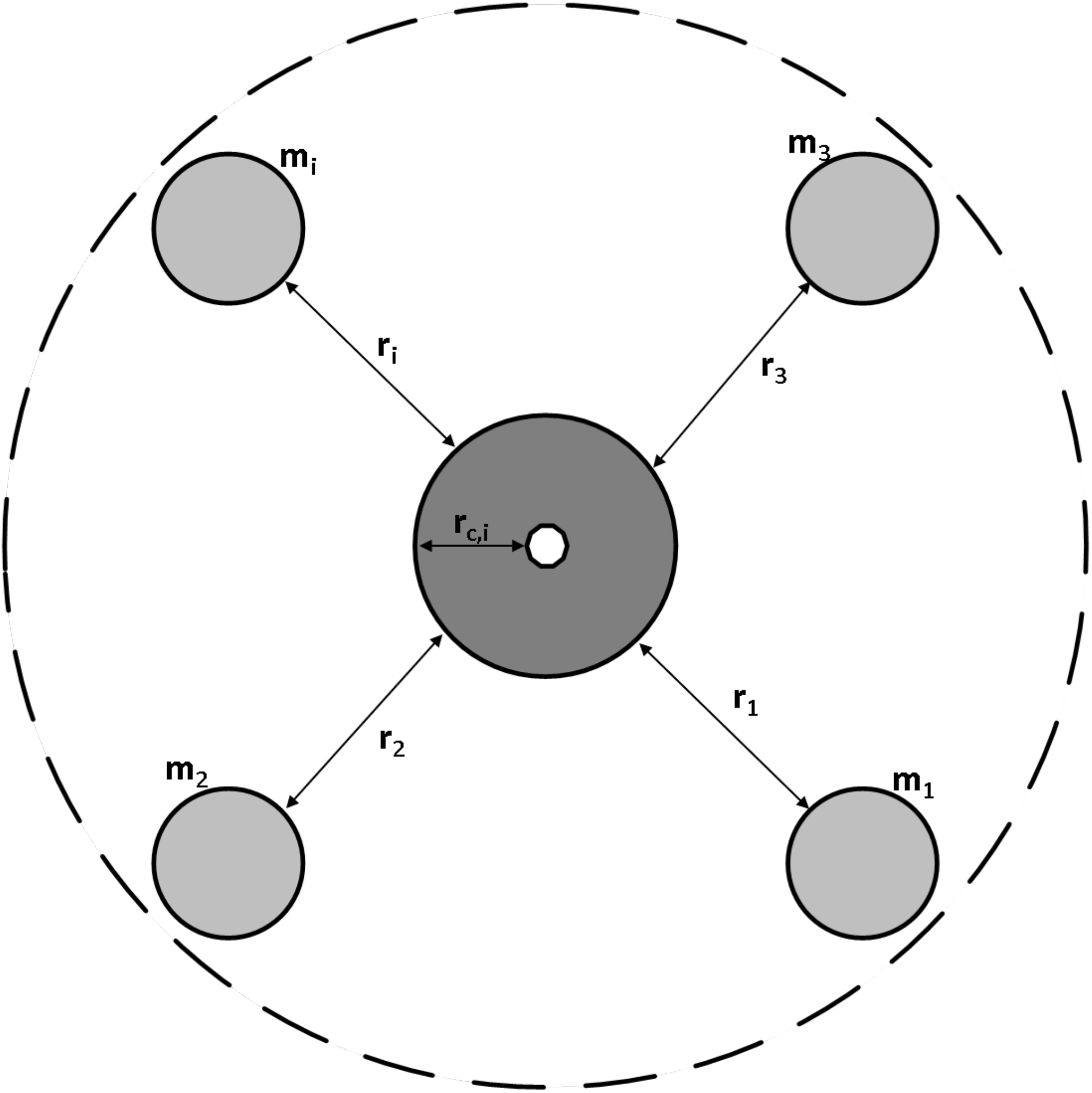}
\caption{Schematic illustration of host stars with masses, $M_{\rm h,\textasteriskcentered}$ and with circumstellar discs of characteristic radius, $r_{\rm c, i}$ in stellar cluster environment. In the stellar cluster, whenever, a perturber of mass, $m_{\rm i} = M_{\rm perturber, \textasteriskcentered}$ visits the host star, the characteristic radius and the mass of the host circumstellar disc is truncated depending on the location and the mass of the perturber. }
\label{fig:1}
\end{figure}

 \subsection{Cluster encounter model}\label{clustermodel}
 To mimic the stellar cluster environments, we subject the calculated disc model to stellar encounters (see the schematic illustration in Figure~\ref{fig:1}) with parameter space typical of the ONC cluster. We assumed one encounter per disc during planet formation history. We also focus on planets that start forming in the discs after encountering the perturber. This assumption is a bare minimum, since it is possible for the host star system to experience multiple stellar encounters during the planet formation history. This assumption was meant to simplify the already intricate aspect of modelling stellar cluster encounters.
 
 The effect of a perturbing star manifests on the host star system in several ways:
 \begin{itemize}
  \item [(i)] Disc property-particularly radius may get truncated by a visiting star in a dense stellar cluster environment. Early numerical simulations \citep{1993MNRAS.261..190C} hint that material could be removed from the host disc by up to one-third of the distance of the encounter. For example, for the closest approach of a perturbing star at 300~au, the disc would be truncated to a radius of 100~au \citep{1996MNRAS.278..303H}. An encounter with the disc could truncates the outer edge and also steepens the density profile of the remaining material in the disc \citep{1997MNRAS.287..148H}, possibly redefining the outskirts of the host disc. The extend of the disc truncation depends on the trajectory of the visiting star relative to the host star. For example, prograde, co-planar encounters are the most destructive to the disc \citep{1998MNRAS.300.1189B}. Material from the disc can also be transferred onto the orbit around the intruding star \citep{2001Icar..153..416K}. Modelling this nature of disc-disc encounters requires inclusion of hydrodynamical effects that comes with intense computational costs. However,  simplified case of only one star being surrounded by a disc is valid, as long as only small fractions of the disc material are captured by the perturbing star \citep{2005A&A...437..967P,2014A&A...565A.130B}.
  
  \item [(ii)] If planetary system already exists around the host star, encounters can disrupt the orbits of the existing planetary systems. For a weak encounter, the planet might remain in orbit around the original parent star. The planet might also coasts along with the perturbing star, and subsequently orbits this star. The planet's trajectory around the perturbing star may be prograde or retrograde, and it is likely to have a high eccentricity and inclination. It is also possible for the planet to be removed from its orbit around the parent star, and eventually becomes a free-floating planet within the star-forming region.
 \end{itemize}
 
 By assuming that discs orbit only the host stars and void in the perturbing star, we modelled effect of stellar encounters on disc properties through the truncation on the disc size. After perturber of mass, $M_{\rm perturber, \textasteriskcentered}$ visits disc-bearing host star of mass, $M_{\rm h,\textasteriskcentered}$, we calculate the final disc size via the  formula presented in \cite{2014A&A...565A.130B} as: 
 
     \begin{equation}\label{eq:rfinal}
          r_{\rm final} = \left\{
     \begin{array}{rl}
       0.28~r_{\rm peri}~{m_{12}}^{-0.32}&:~{\rm for}~ r_{\rm final} \leq r_{\rm c} \\
        r_{\rm c}&:{\rm otherwise}
     \end{array}.
   \right.
    \end{equation}

$m_{12}~{\rm and}~ r_{\rm peri} $ are the mass ratio between the encountering and the disc-bearing star and the periastron distance, respectively. Periastron distances, $r_{\rm peri}$ is calculated from the periastron ratio distribution typical of the ONC. We describe the periastron distribution in the later section~\ref{Initcond}. Because discs might spread viscously, equation~\ref{eq:rfinal} might deviate for a viscous disc. Therefore we used throughout this study moderately low viscosity values, in particular, the viscosity-parameter, $\alpha_{\rm visc}$ was fixed to $10^{-3}$ and $6\times 10^{-4}$\footnote{ Such low viscosity-parameter have been found to significantly damp type~II migration rates of forming gas giant planets, possibly saving them from being engulfed by their host star \citep{2021MNRAS.501.2017N}}. As in \cite{2014A&A...565A.130B},  the disc truncation model adopted here holds for the most destructive encounters (i.e, the prograde, the coplanar, and the parabolic (e = 1)). Therefore, the resulting disc sizes are equivalent to the maximal impact of stellar encounters and represents the floor limits for the disc sizes compared to the encounters on inclined and/or hyperbolic orbits \citep[e.g,][]{1993MNRAS.261..190C,2005ApJ...629..526P}.

 Whenever there is stellar encounter in our setup, $r_{\rm c}$ changes. We modelled the shrinkage in $r_{\rm c}$ due to stellar flyby to lead to mass loss just exterior to the point of disc truncation, but, without significant change in mass distribution interior to the point of the disc truncation. We think with this assumption, the prior and post flyby disc profile nearly remains the same, but differ with a drop in surface density at the location of the truncation as in  \cite{2014A&A...565A.130B}. It is this truncation location that sets the updated final size of our disc profile. We calculate the remaining mass of the disc, $M_{\rm disc, rem}$ as:
\begin{equation}\label{eq:mdiscfinal}
M_{\rm disc, rem} = \left(\frac{r_{\rm final}}{r_{\rm c}}\right)^{2}{M_{\rm disc}} .
\end{equation}
 It is from these final disc sizes and disc masses of the post-encounter environment, we calculate the initial gas and dust surface density. We at later stage evolve the gas and dust surface density while taking into account the realistic grain population distribution in the discs. 


\subsection{Grain size distribution}\label{grainpop}

Small micrometer sized dust grains can grow in protoplanetary discs through coagulation \citep{2008A&A...480..859B, 2011A&A...525A..11B} and condensation \citep{2013A&A...552A.137R}. The growth of the particles is limited by fragmentation of the grains, which arises when the relative velocities of the individual grains become so large that the grains fragment. The laboratory measurement of the threshold speed (fragmentation velocity, $u_{\rm f}$) suggest values around $1-10$~m/s \citep{2015ApJ...798...34G}. The maximum grain size dust particles can reach is thus given by
 \begin{equation}
  \label{eq:smax}
  s_{\rm max} \approx \frac{2 \Sigma_{\rm g} u_{\rm f}^2}{\pi \alpha \rho_{\rm s} c_{\rm s}^2} \ .
 \end{equation} 
 $\rho_{\rm s}$ is the density of the pebbles, assumed to be constant and set to 1.6~g/cm$^3$, and $c_{\rm s}$ is the sound speed of the gas.

The exact shape of the grain size distribution is a complex interplay between settling, coagulation, cratering and fragmentation. 
We follow the two-population dust evolution model of \citet{2012A&A...539A.148B} from which we reconstruct full grain size distribution using the reconstruction model of \cite{2015ApJ...813L..14B}.

The maximal grain size is quadratically proportional to the fragmentation velocity of the grains and inversely proportional to the $\alpha$ parameter (eq.~\ref{eq:smax}). We remind the reader that the grain size distributions in our simulations feature a fixed combination of $\alpha=10^{-3}; u_{\rm f}=10~{\rm m/s}$ and $\alpha=6\times10^{-4}; u_{\rm f}=10~{\rm m/s}$. As hinted before, grain size distribution is sensitive to at least $\alpha~{\rm and}~u_{\rm f}$. The fixed choice for viscosity-parameter and the fragmentation velocity in the simulations is a simplification also made in \citep[e.g.][]{2021A&A...645A.131V,2021A&A...645A.132V,2021A&A...647A..15D,2021arXiv210513267S,2021arXiv210903589S} that built on the two-population dust evolution model of \citet{2012A&A...539A.148B}. 

In principle, the grains interact with the gas in the protoplanetary disc, which suck from them angular momentum, resulting in an inward drift of pebbles \citep{1977MNRAS.180...57W}. For simplicity reasons, we consider only the Epstein drag, not the Stokes regime. The drift rate of particles depends on their Stokes number, defined as 
\begin{equation}
 \label{eq:Stokes}
 St = \frac{\pi}{2} \frac{\rho_{\rm s} a}{\Sigma_{\rm g}} \ .
\end{equation}
The drift can lead to a pile-up of grains in the inner regions of the disc \citep{2012A&A...539A.148B}. As in  \cite{2012A&A...539A.148B}, our work took into account the balancing between grain growth, fragmentation and drift size limits.

\subsection{Planet formation model}

The planetary seeds in this study are introduced at pebble transition mass $0.01$~$M_{\rm E}$ at various distances $a_0$ from the central star, at various initial starting times $t_0$  and in discs with varying lifetimes $t_{\rm life}$. Planetary cores initially accrete material via the inefficient 3D pebble 
accretion branch \citep{2012A&A...544A..32L} until they become massive enough such that their Hill radius exceeds the pebble scale height  ($R_{\rm H}>H_{\rm peb}$, see \cite{2015Icar..258..418M} for more discussion). During solid accretion regime, planets migrate in the type~I migration regime that scales linearly with the mass of the growing embryo. The type~I migration rate is calculated using the torque prescription of \cite{2011MNRAS.410..293P}. At low $\alpha$ values, type~I migration is increasingly fast with increasing planetary mass due to early saturation of corotation torques. However, as numerically experimented in \cite{2021MNRAS.501.2017N}, the speed of Type~I migration can be  reduced if we consider the dynamical corotation torque \citep{2014MNRAS.444.2031P}. The dynamical corotation torque was therefore incorporated as in \cite{2021MNRAS.501.2017N} and is particularly effective in reducing type~I migration in low-viscosity discs with shallow surface density profile, like the one assumed here. 
 
 Planetary cores accrete pebbles, until it reaches pebble isolation mass. At pebble isolation, the pressure bump generated by the planet block the accretion of pebble, instead, the cores start accreting gas. We modelled the pebble isolation mass via the criterion that allow diffusion of particles across the pressure bump \citep{2018A&A...612A..30B}. We remind the reader that the pebble accretion scheme in this work follows the concurrent accretion of multiple pebble sizes \citep{2022MNRAS.510.1298A}, with pebble size distributions derived from the full grain size distribution (see sub section~\ref{grainpop}).
 
 During the gas accretion phase, we captured the contribution of the envelope opacity ($\kappa_{\rm env}$) to the planetary formation model by focusing on it's influence on the envelope contraction rates\footnote{In principle, opacity influence are not only reflected in gas envelope contraction rates but can also affect migration rates of planetary cores \citep{2020A&A...640A..63S,2021A&A...647A..96B}. Therefore our results hold when the contribution of opacity on migration rates are minimal.}, where we use a uniform envelope opacity of $\kappa_{\rm env}=0.05~\rm cm^{2}g^{-1}$. We acknowledge that planetary envelope opacity varies with grain sizes. Our envelope opacity choice is a simplification also made in many planet formation simulations \citep[e.g,][]{2015A&A...582A.112B}. In addition, our envelope opacity choice is also in agreement with the study by \cite{2008Icar..194..368M}, who found that the dust grain opacity in most of the radiative zone of the planet's envelope is of the order of $10^{-2}$. This is also in agreement with \cite{2015IJAsB..14..201M}. The gas accretion approach in this work follows the detailed gas accretion prescriptions in \cite{2021MNRAS.501.2017N}, where envelope contraction rate was modelled via \cite{2000ApJ...537.1013I}. In \cite{2000ApJ...537.1013I}, envelope contraction rates scale with planetary core mass and the envelope opacity, $\kappa_{\rm env}$.

We stop the simulation when the planet either (i) reaches 0.04~au, where we assume migrating planets are trapped at the inner disk edge \citep{2006ApJ...642..478M,2019A&A...630A.147F} or (ii) at the time of disc dispersion for the planets that do not reach the inner edge.

\section{Initial Conditions}\label{Initcond}
Planet population syntheses require sampling important disc and planet formation parameters as initial inputs. Although the existing initial conditions distributions of the available global planet formation models may miss essential physics that follows their assumed initial distributions \citep{2022arXiv220201500V}, useful planet formation insights can still be extracted from such initial distributions that do not capture the dynamical evolution of the planetary initial starting distributions. For simplicity reasons, we did  not mostly modelled (as in the most global planet formation models) the evolution of initial conditions, but rather assumed mostly fixed ones that follow either educated guesses or distributions from observational surveys. 

The importance of stellar encounter in this work is reflected in setting the disc-properties-particularly the size and the mass. With the initial disc mass and the host star mass set to $0.1~M_{\bigodot}$ and $1~M_{\bigodot}$, respectively, we sampled the relevant stellar encounter parameters as follows:

\begin{itemize}
 \item [--] The stellar masses of the perturbers are randomly drawn from \cite{2002MNRAS.336.1188K} mass distribution with mass range fixed to stellar mass limits of the ONC [$0.08~M_{\bigodot},~40~M_{\bigodot}$]. It is from this perturber mass distribution that we calculate the perturber mass to host star mass ratios.
 \item [--] The periastron ratio of the encounter follows the periastra ratio limits of the ONC parameter space and with the distribution that mimics linear and logarithmic distributions.
\end{itemize}

Besides the configurations mentioned below, we assumed the planet formation initial conditions as in \cite{2018MNRAS.474..886N,2019MNRAS.488.3625N}:
\begin{itemize}
\item  [--] Cores implanted at pebble transition mass
\item  [--] Cores implanted in the range [0.1~$au$: 0.85~$r_{\rm final}~au$]
\item  [--] linear distribution of the initial core injection times in the range [0.1:1]~My.
 \item [--] The disc lifetime distribution follows the \cite{2009AIPC.1158....3M} survey, where there is exponentially decaying for increasing lifetimes with a half lifetime of 2.5~My, where we fixed the minimal and maximal lifetime of the disc to 0.1 and 10~My, respectively.
 \item [--] The dust-to-gas ratio, $Z$ was computed as:

\begin{equation}
 Z= 0.0179\times 10^{\rm [Fe/H]}.
\end{equation} We followed the computation in \cite{2021arXiv210513267S}, where $\rm [Fe/H] = 0$ corresponds to 0.0179. The metallicity, ${\rm [Fe/H]}$ of the disc follows a Gaussian distribution centered around 0.0179.
\end{itemize}

\begin{figure}
\includegraphics[width=\linewidth]{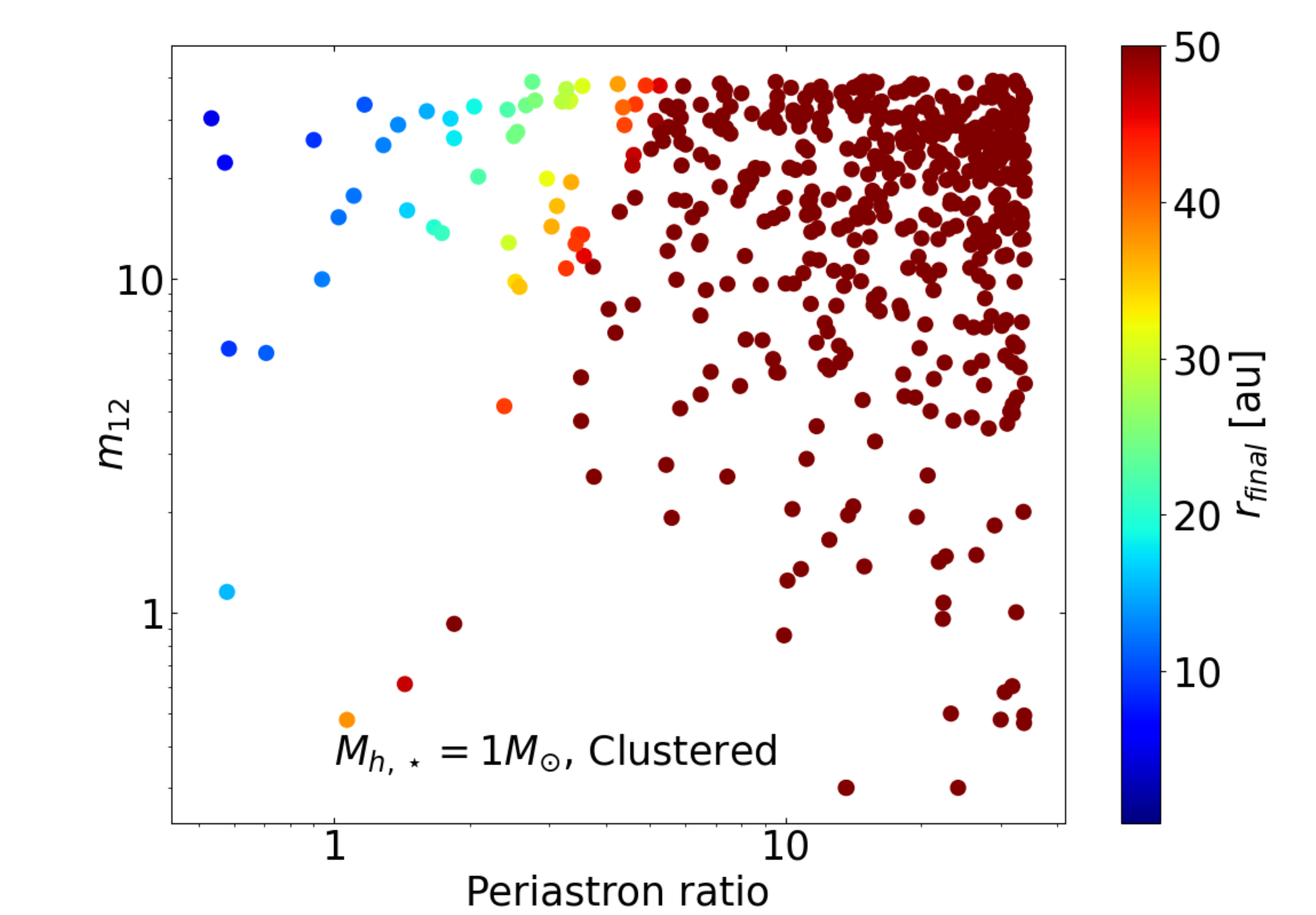}
\caption{The final disc size, $r_{\rm final}$ as a function of perturbers' periastron ratios and mass ratios, $m_{12}$. The plot features linear distribution in perturbers' periastron ratios and mass ratios with host star mass fixed to $1~M_{\bigodot}$. The multiple colors show the remaining disc size after disc truncation by stellar encounters.}
\label{fig:2mratio}
\end{figure}

\begin{figure}
\includegraphics[width=\linewidth]{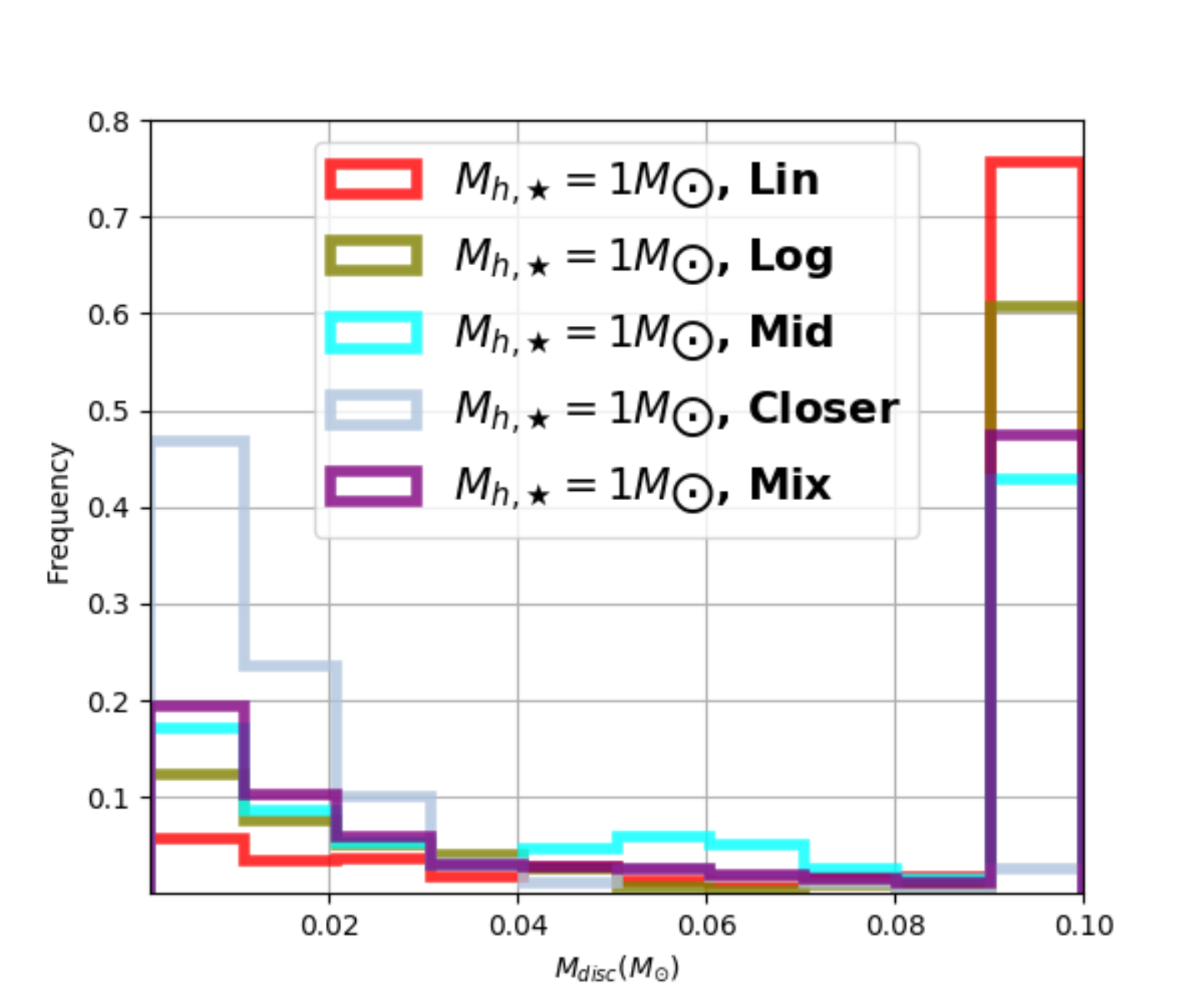}
\caption{Post encounter host disc mass distributions for the different distributions in periastron ratios; linear distribution (the red curve that correspond to the final disc sizes shown in Figure~\ref{fig:2mratio}), logarithmic distribution (olive curve), mid distribution (cyan curve), closer distribution (light blue curve) and mixed distribution (purple curve).}
\label{fig:2paramImpact}
\end{figure}

\begin{figure}
        \centering
        \includegraphics[width=\columnwidth,height=2.45in]{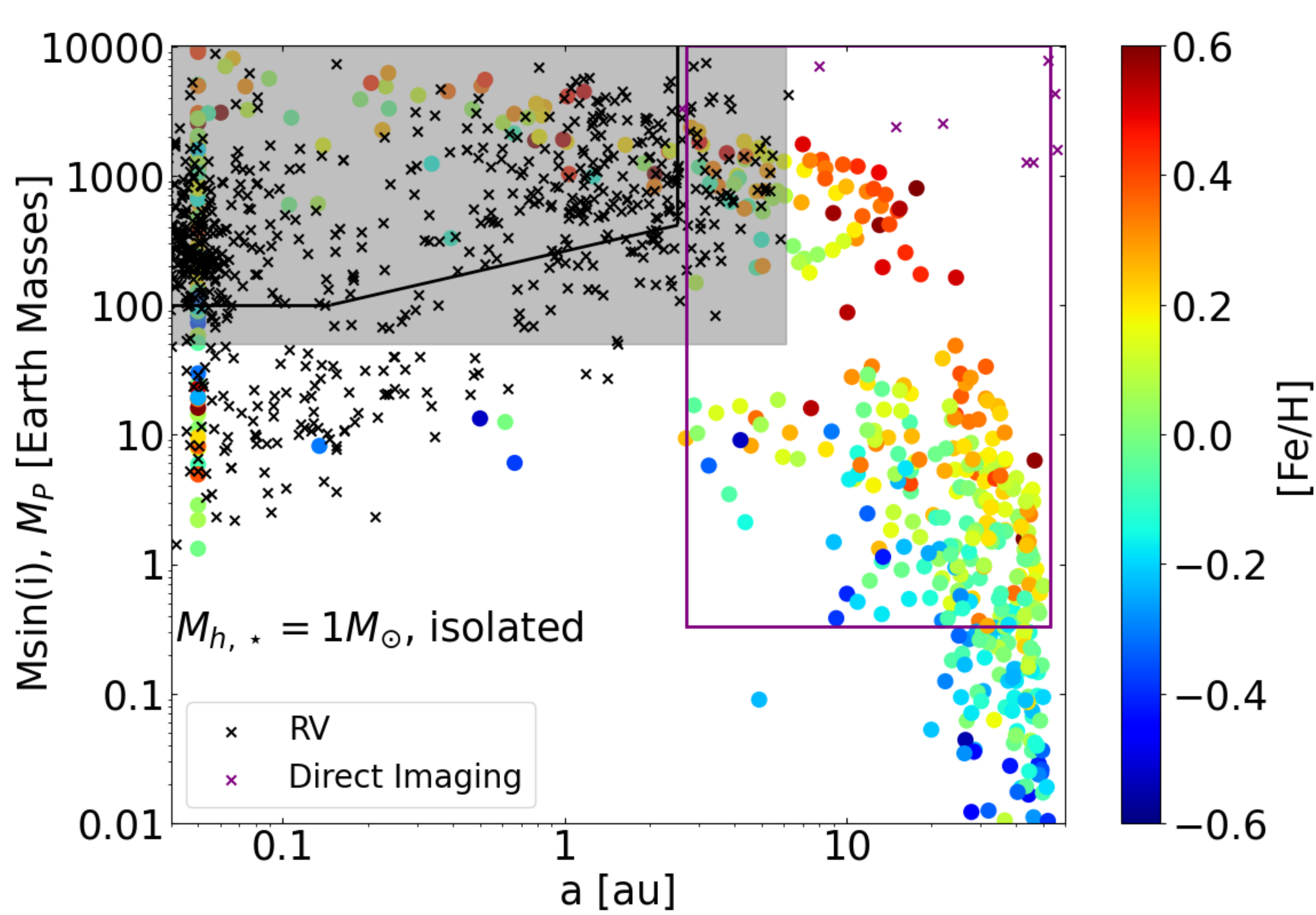}
        \includegraphics[width=\columnwidth,height=2.45in]{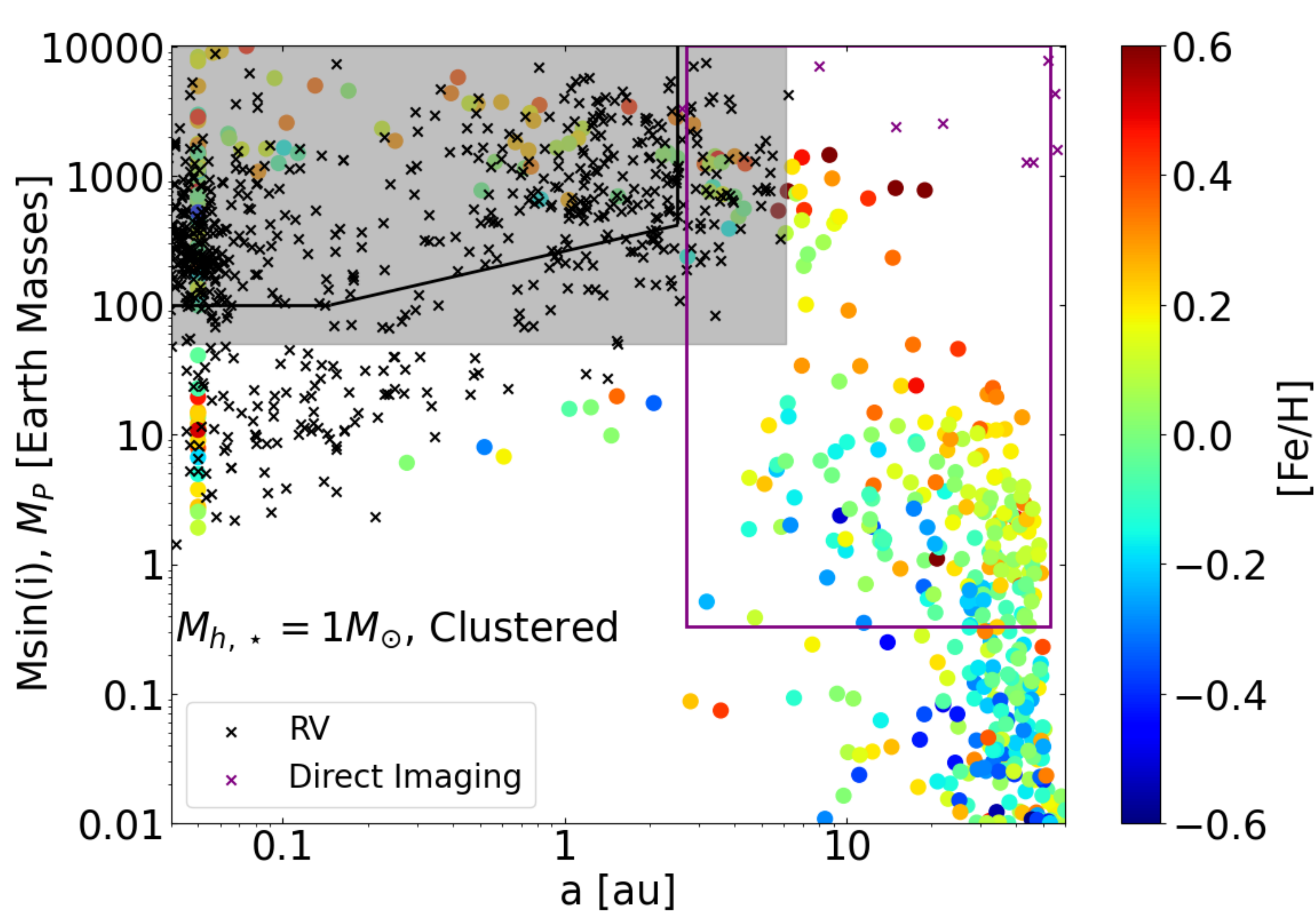}
        \includegraphics[width=\columnwidth,height=2.45in]{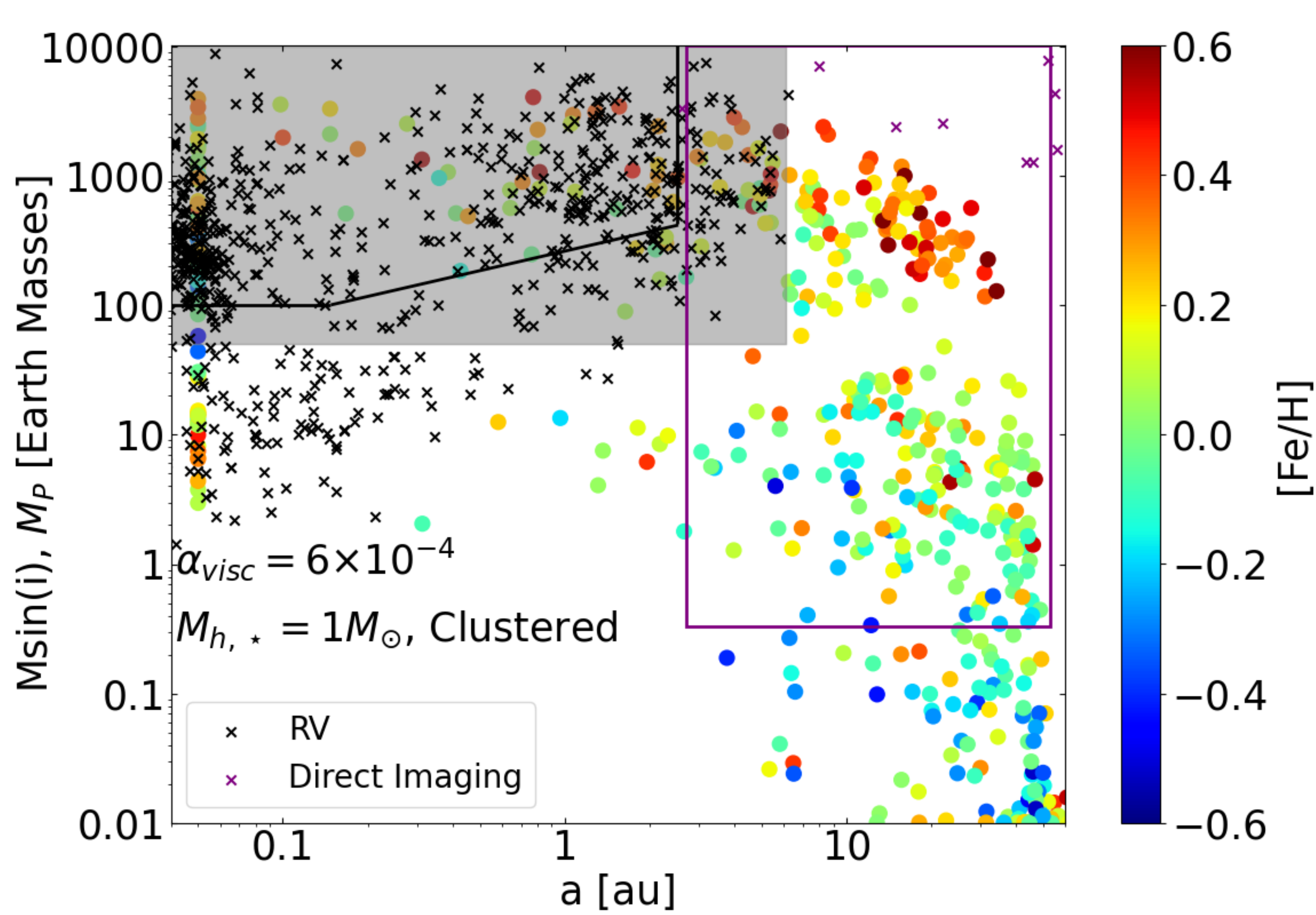}
\caption{ Synthesis plots showing the final planetary masses and positions. The multicolored dots show the synthesized planets as a function of disc metallicity. The black crosses and the purple crosses represent RV exoplanet data from exoplanet.org repository and the directly imaged exoplanets. Planets inside the purple rectangle show the detection limits of the microlensing survey of \protect\cite{2016ApJ...833..145S}. $Msin(i)$ is the RV planet mass and $M_{\rm P}$ comes from the simulations. Planets inside the top left quadrant marked by the black lines represent giant planets that were observed until completeness by the RV survey of \protect\cite{2010PASP..122..905J}. We show as grey shade, planets with $M_{\rm P}~>~50$~Earth masses that we compare to the semi-major axis distribution originating from \protect\cite{2011arXiv1109.2497M} in Figure~\ref{fig:5}. The top  plot shows the simulation without disc truncation. The middle and the bottom plots feature simulations with disc truncation, but, with ${\alpha}_{\rm visc}$ set to $10^{-3}$ and $6{\times}10^{-4}$, respectively.}
\label{fig:3}
\end{figure}

\begin{figure*}
        \centering
        \includegraphics[width=8.5in,height=5.5in]{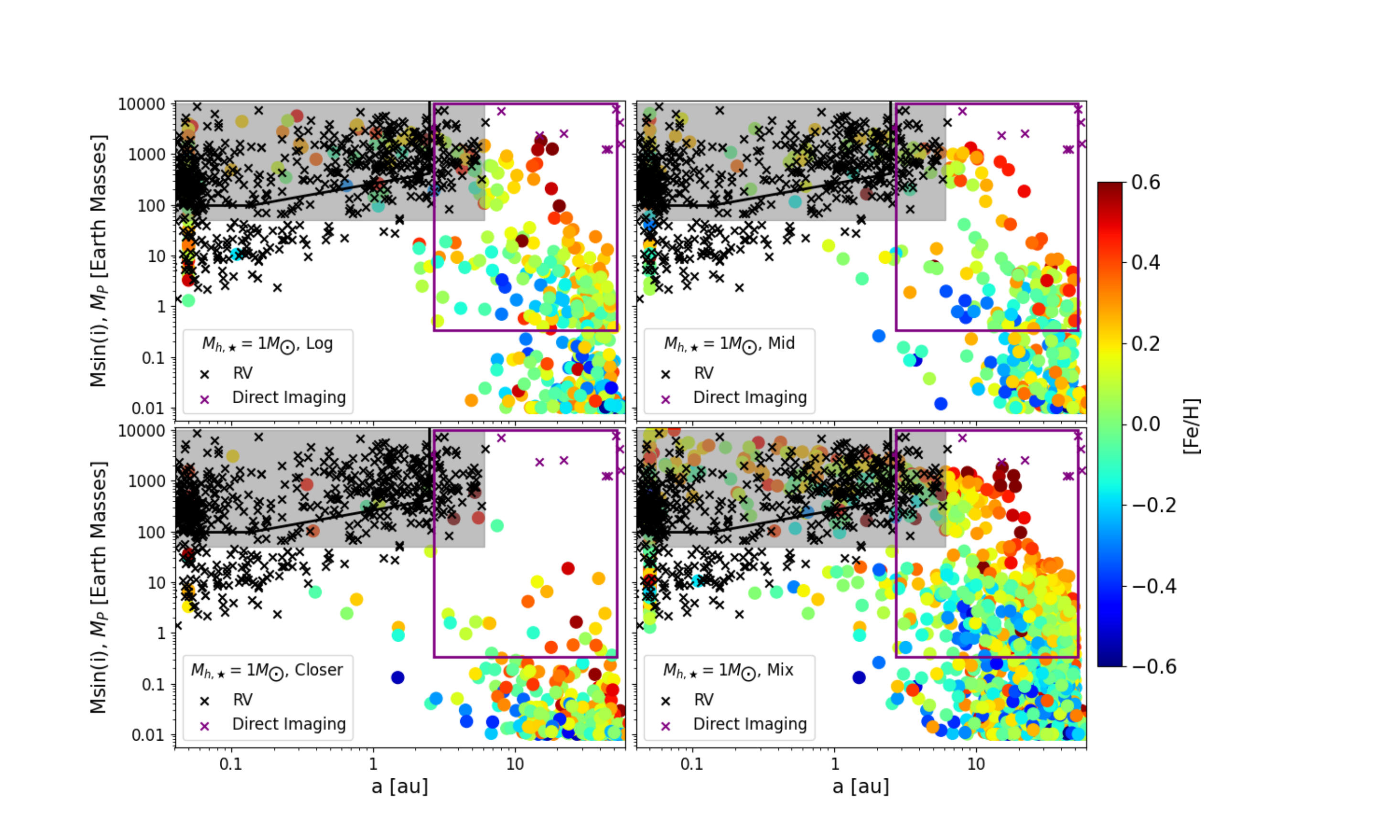}

  \caption{Synthesis plots showing the impact of periastron ratio distributions on disc truncation and consequently on the final planetary masses and orbital distances. The top plots features logarithmic distribution in periastron ratios (left panel) and mid distribution in periastron ratios (right panel). The bottom plots features closer distribution in periastron ratios (left panel) and mixed distribution in periastron ratios (right panel). Apart from that, the lines and dots have  the same meaning as in Figure~\ref{fig:3}.}
         \label{fig:4}
 \end{figure*}

\section{Truncation of host disc masses by stellar encounters}\label{discmass}
Because the stellar encounter model considered in this work operates by changing the disc properties-particularly the disc size, $r_{\rm c}$ and the disc mass, $M_{\rm disc}$, we begin by performing parametric experiment on the stellar encounter parameters that give optimal effect on the disc-properties (the disc size and the disc mass). Figure~\ref{fig:2mratio} shows within the assumed stellar encounter model, disc truncation by visiting stars are much more dependent on the locations of the perturbers (Periastron ratio) than the masses of the perturbers (mass ratio, $m_{12}$). It is also noticeable from Figure~\ref{fig:2mratio} that distant perturbers have very minimal impact on the discs' final sizes, unless the perturbers are relatively very massive in comparison to the host star mass. For a given extent of disc-size truncation, our model calculated the truncated disc mass, following equation~\ref{eq:mdiscfinal}. The impact of stellar encounter in our disc properties is therefore reflected through the reduction in the disc's size and the disc's mass. Since the disc truncation in our cluster models depend strongly on the periastron ratio and weakly on the mass ratios (see Figure~\ref{fig:2mratio}), we perform parametric experiment with periastron ratio as an input and show the result in Figure~\ref{fig:2paramImpact}. Figure~\ref{fig:2paramImpact} shows the distributions of the resultant host disc masses after visit by perturbers whose mass ratios followed linear distribution. 

Although the periastron ratio limits in ONC clusters are known, little is known about the exact distribution shapes of the periastron ratios. We therefore performed parametric experiment within periastron ratio limit of ONC by assuming not only linear and logarithmic distributions in periastron ratio, but also, a closer, middle and mixed distributions in periastron ratios. The closer distribution feature linear distribution inside periastron ratio of 5. Meanwhile the middle distribution relate to the linear distribution inside of periastron ratio of 15. The mixed distribution features all the aforementioned distributions; linear, logarithmic, closed and the middle distribution. It is clear from Figure~\ref{fig:2paramImpact} that the maximal impact on the host disc mass occurs when the perturbers are closer to the host star (see the light blue host disc mass distribution in Figure~\ref{fig:2paramImpact}) with the disc mass following an exponential decaying function. Host disc mass relates to the available building blocks for forming planets. For a fixed disc-parameters, a disc with high disc mass readily forms massive planets compared to the low disc mass. Truncation reduces the available host disc mass, consequently impacting planet formation. Disc truncation also shift the planet formation regions closer to their host stars than before the encounters. Therefore, planetary seeds grow in the post-encounter host discs at closer radial locations that might for fast-inwards migration drift very fast to the inner-edge of the disc. We show the impact of these host disc truncation on global planet formations in Figure~\ref{fig:3} and Figure~\ref{fig:4}.

\section{The role of stellar encounters in shaping planet population}\label{popsynth}
In this section, we show the results of planet population synthesis simulations with and without stellar encounters.
Planet population synthesis in this work calculates planetary growth and migration from small embryos, interlacing pebble accretion, gas accretion onto the planet, type I and II migrations, and disc truncation.

Figure~\ref{fig:3} and Figure~\ref{fig:4} show the results of planet population synthesis simulations of 500 systems around solar-type stars, with host disc masses fixed to $0.1M_{\bigodot}$.  In general Figure~\ref{fig:3} and Figure~\ref{fig:4} revealed that through stellar encounters, formation of gas giant planets is less supported. This is due to the disc truncation that reduces the disc mass. Figure~\ref{fig:3} probes the impact of disc truncation and viscosity on the distributions of planetary population using linear distribution in periastron ratio. Because with linear distributions in periastron ratio the disc is not substantially truncated, the post encounter disc mass distributions mostly reside at $0.1 M_{\bigodot}$ (see the red curve in Figure~\ref{fig:2paramImpact}). There is therefore enough material to form relatively massive planets, even in the microlensing field of view (see the middle plot in Figure~\ref{fig:3}). We show in bottom plot of Figure~\ref{fig:3} how viscosity impact the contributions of stellar encounter on planetary population by assuming as in \cite{2021MNRAS.501.2017N} a lower viscosity parameter, ${\alpha}_{\rm visc}=6\times10^{-4}$. With this viscosity parameter, there is clearly less massive giant planets and more gas giant planets trapped in the microlensing field of view (see bottom plot in Figure~\ref{fig:3}). This hints that forming the cold gas giant in post encounter discs might require lower viscosities and inward migration speeds.

Figure~\ref{fig:4} shows the impact of the different periastron ratio parameter configurations on planet population. As revealed in Figure~\ref{fig:2paramImpact}, a closely passing perturbing stars substantially truncate the host disc and therefore limits the amount of available building blocks for forming gas giant planets. In such truncated discs, planet formation regions are also limited to regions close to the host stars and forming planets could potentially undergo fast inward migration and engulfed by the host star.

Using the relative proportions between the different categories of synthesized planets, we show how disc truncation by stellar encounters impact the synthesized sub-populations of planets similar to the basic statistical operations performed in \cite{2018MNRAS.474..886N} (see Table~\ref{tab:1}). As in \cite{2018MNRAS.474..886N}, the  isolated field simulations show that the gas giant planets are the majority compared to the super Earth planets. The simulation particularly show 18.2 \% in super Earths, 7.6 \% in Neptunian planets, 36.2 \% in gas giant planets (hot, warm or cold Jupiters) and  38 \% in low mass planets.  This is contrary to exoplanet surveys that show; that gas giant planets are rare and form mostly at high metallicity \citep{2005ApJ...622.1102F,2010PASP..122..905J}. The simulations overly predict gas giant planets (hot, warm or cold Jupiters) than the super Earth planets. This presents another nexus of conflict with exoplanet surveys that show nearly 50 \% of observed exoplanets are rocky planets within 1~au \citep{2005ApJ...622.1102F,2013ApJ...766...81F}. Including disc truncation from stellar encounters within linear periastron ratio distribution, overall promotes the formation of low mass planets and reduces the proportion of gas giant planets to 30.2 \%. Past studies, point to the same conclusion, where it was shown disc truncation limits the disc's total size and consequently planetary frequency \citep[e.g,][]{2001MNRAS.325..449S,2012ApJ...756..123O,2014MNRAS.441.2094R,2014A&A...565A..32S,2016MNRAS.457..313P,2016ApJ...828...48V,2019MNRAS.482..732C,2021arXiv210107826C}.

The nominal simulation without stellar encounter shows that the cold Jupiters (16.2 \%) are the majority in comparison to the hot Jupiters (15.6 \%), consistent with observation that shows cold Jupiters are more frequent than hot Jupiters. This statistics however changes on inclusion of stellar encounter, with the cold Jupiters and the hot Jupiters now taking a proportion of 9.6 \% and 12.2 \%, respectively (see the data in Table~\ref{tab:1} for the linear distribution in periastron ratios). We attribute, the overall reduction in gas giant planets in stellar clusters to the disc truncation that reduces the overall disc mass available for building the gas giant planets. However, in lower viscosity environment of $\alpha_{\rm visc}=6\times 10^{-4}$, the cold Jupiters (25.4 \%) are much more common than the hot Jupiters (9 \%). This is because, in low viscosity environment planetary migration is slow, with planets consequently residing close to the outskirts of the natal disc.

The ratio of hot to cold Jupiters changes with changing periastron ratio, for distributions other than linear distribution (logarithmic, mid, closer and mixed periastron). For example, the ratio of hot to cold Jupiters is less than a unity for distributions other than linear (see the part of Table~\ref{tab:1} that features logarithmic, mid, closer and mixed distributions in the periastron ratio distributions). With these distributions in periastron ratios, the stellar encounter hinders gas giant planet formation more compared to the encounter model with linear periastron ratio distribution. For example, the gas giant planet proportion is 21 \%, 25.4 \%, 4.3 \% and 18 \% for cluster model with logarithmic, mid, closer and mixed periastron ratio distribution. These values of gas giant planet proportions are all low in comparison to the 36.2 \% of gas giant planet synthesized via the stellar encounter with linear distribution in periastron ratios. This implies with logarithmic, mid, closer and mixed distribution patterns in periastron ratio, impact of disc truncation on gas giant planet population reduction is substantial, with the reduction much more pronounced for stellar encounter model that follows closer periastron ratio configuration.

 
In the next section we discuss how the various simulations compare with the debiased semi major axis data from \cite{2011arXiv1109.2497M}, the metallicity correlation data from \cite{2010PASP..122..905J} and the microlensing data from \cite{2016ApJ...833..145S}. 
  \begin{table}
\caption{Fractions of synthesised planets. SE and NP shows super Earth planet ($2 {\rm M}_{\rm E} < M_{\rm P} < 16 {\rm M}_{\rm E}$ ), Neptunian ($ 16 {\rm M}_{\rm E} < M_{\rm P} < 100 {\rm M}_{\rm E}$ ), respectively. HJ, WJ, and CJ represents hot Jupiters ($r_{\rm p} < 0.1$ AU), warm Jupiters (0.1 AU $< r_{\rm p}<$ 1.0 AU) and cold Jupiters ($r_{\rm p} >$ 1.0 AU), where all Jupiter type planets have $ M_{\rm P} > 100 {\rm M}_{\rm E}$. The population statistics shows  simulation result shown in Figure~\ref{fig:3} and \ref{fig:4}. The table shows the fractions of the relevant planetary category. The remaining fractions for each simulation depicts, the proportions of failed cores ($M_{\rm P}<2 {\rm M}_{\rm E}$) in our simulations. $M_{\rm disc}$ and Env are the host disc mass and the disc environment of the simulation, respectively.}
\label{tab:1}
\begin{center}
Linear distribution
\footnotesize
\begin{tabular}{lcccccccccc}

$r_{\rm peri}$ & Env & SE &NP &HJ&WJ&CJ&\\ 
               &  &  &  &  &  &  &\\ 
              \hline        
         NA&Isolated,  $\alpha_{\rm visc}=1\times10^{-3}$&0.182&0.08&0.156&0.044&0.162&\\
        Lin&Clustered,  $\alpha_{\rm visc}=1\times10^{-3}$&0.196&0.038&0.122&0.046&0.096&\\
        Lin&Clustered, $\alpha_{\rm visc}=6\times10^{-4}$&0.224&0.064&0.090&0.042&0.254&\\
		\hline	       
\end{tabular}\\

 Logarithmic, Mid, Closer and Mixed distribution        

 \begin{tabular}{lcccccccccc}

$r_{\rm peri}$ & Env & SE &NP &HJ&WJ&CJ&\\ 
               &  &  &  &  &  &  &\\ 
              \hline        
         Log&Clustered,  $\alpha_{\rm visc}=1\times10^{-3}$&0.166&0.058&0.072&0.032&0.106&\\
        Mid&Clustered,  $\alpha_{\rm visc}=1\times10^{-3}$&0.140&0.033&0.065&0.093&0.096&\\
        Closer&Clustered, $\alpha_{\rm visc}=1\times10^{-3}$&0.029&0.010&0.010&0.019&0.014&\\
        Mixed&Clustered, $\alpha_{\rm visc}=1\times10^{-3}$&0.137&0.036&0.070&0.030&0.080&\\

	\hline		       
\end{tabular}

\end{center}
\end{table}
 \section{comparison to the observed extrasolar population}\label{comparison}
 
%

   \begin{figure*}
        \centering
         \includegraphics[width=\columnwidth, 
height=2.35in]{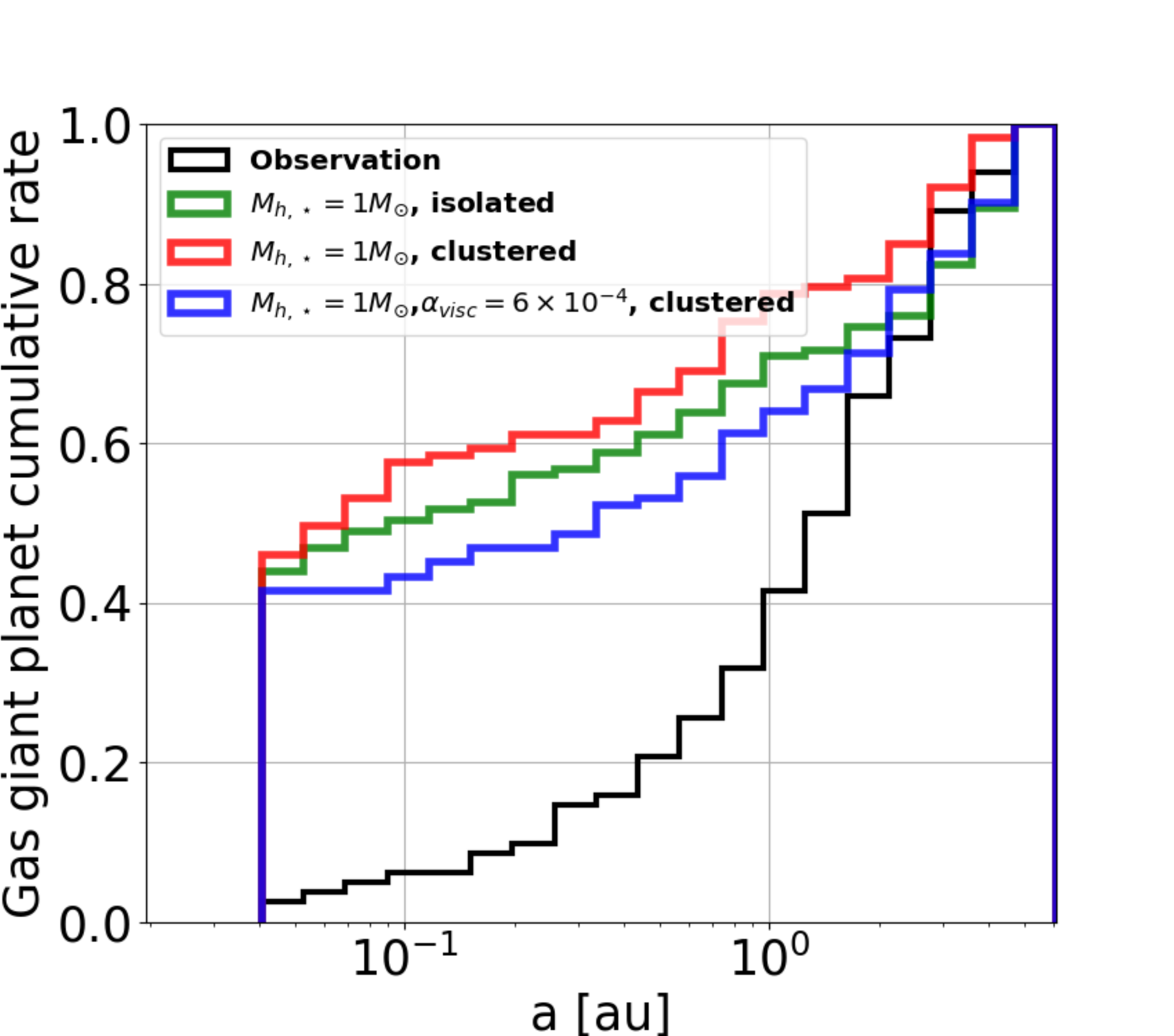}\quad
           \hfill
        \includegraphics[width=\columnwidth,height=2.35in]{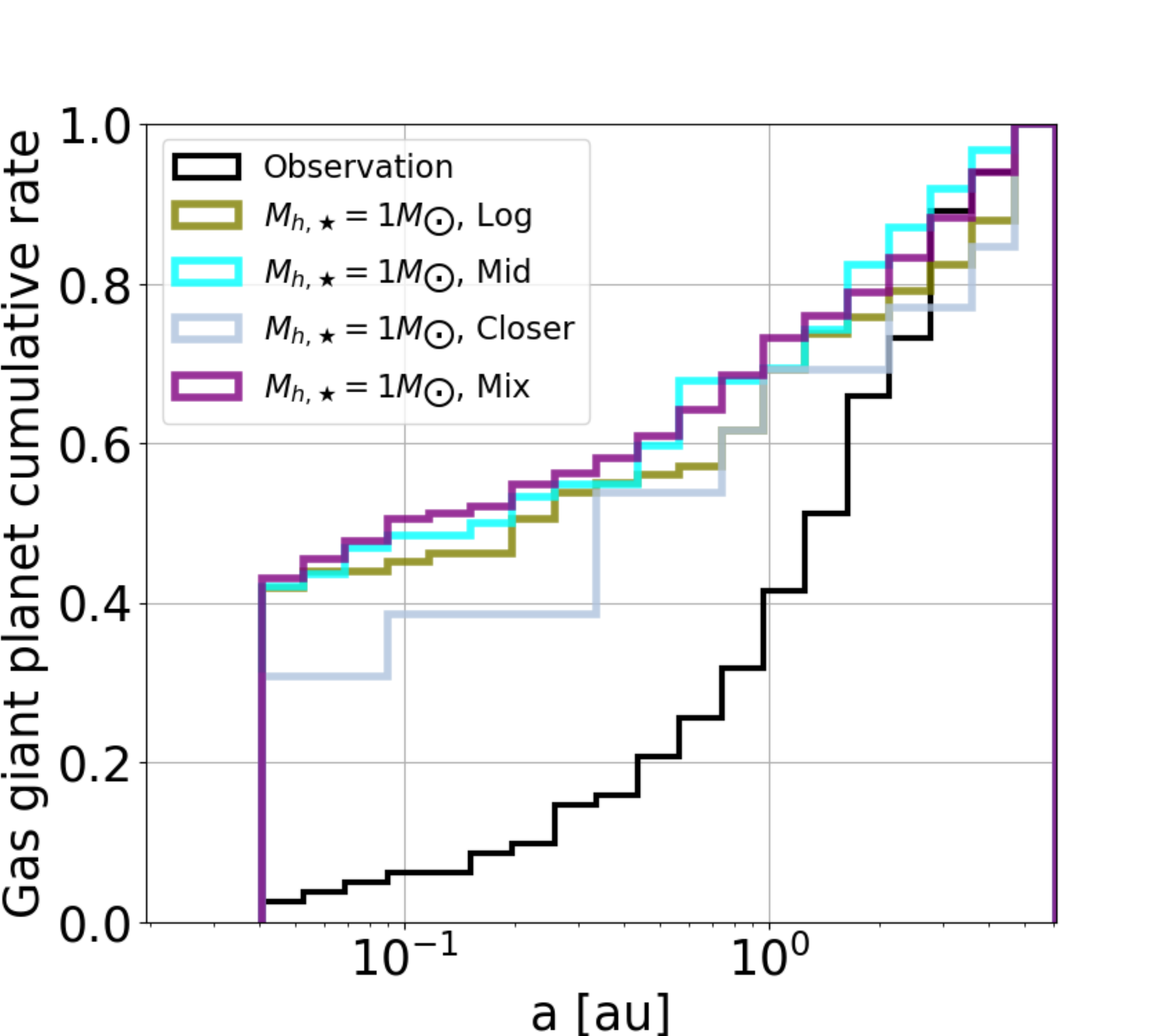} 

 \caption{Cumulative distribution of the planetary semi-major axis $a$ for planets with $Msin(i)~>~50$~Earth mass, marked by the gray box in Figure~\ref{fig:3} \&~\ref{fig:4}. The black curve represents the cumulative rate derived from the debiased data of \protect\cite{2011arXiv1109.2497M}. The remaining colored curves represent the models from our simulation that features the different stellar and disc environment configurations. The left and the right panel plots show the occurrence rates with linear distribution and  other periastron ratio distribution (logarithmic, mid, close and mixed distributions), respectively. The occurrence rate for gaseous giant planets is strongly increasing with the logarithm of $a$. }
         \label{fig:5}
 \end{figure*}

\begin{figure*}
        \centering
         \includegraphics[width=\columnwidth, 
height=2.35in]{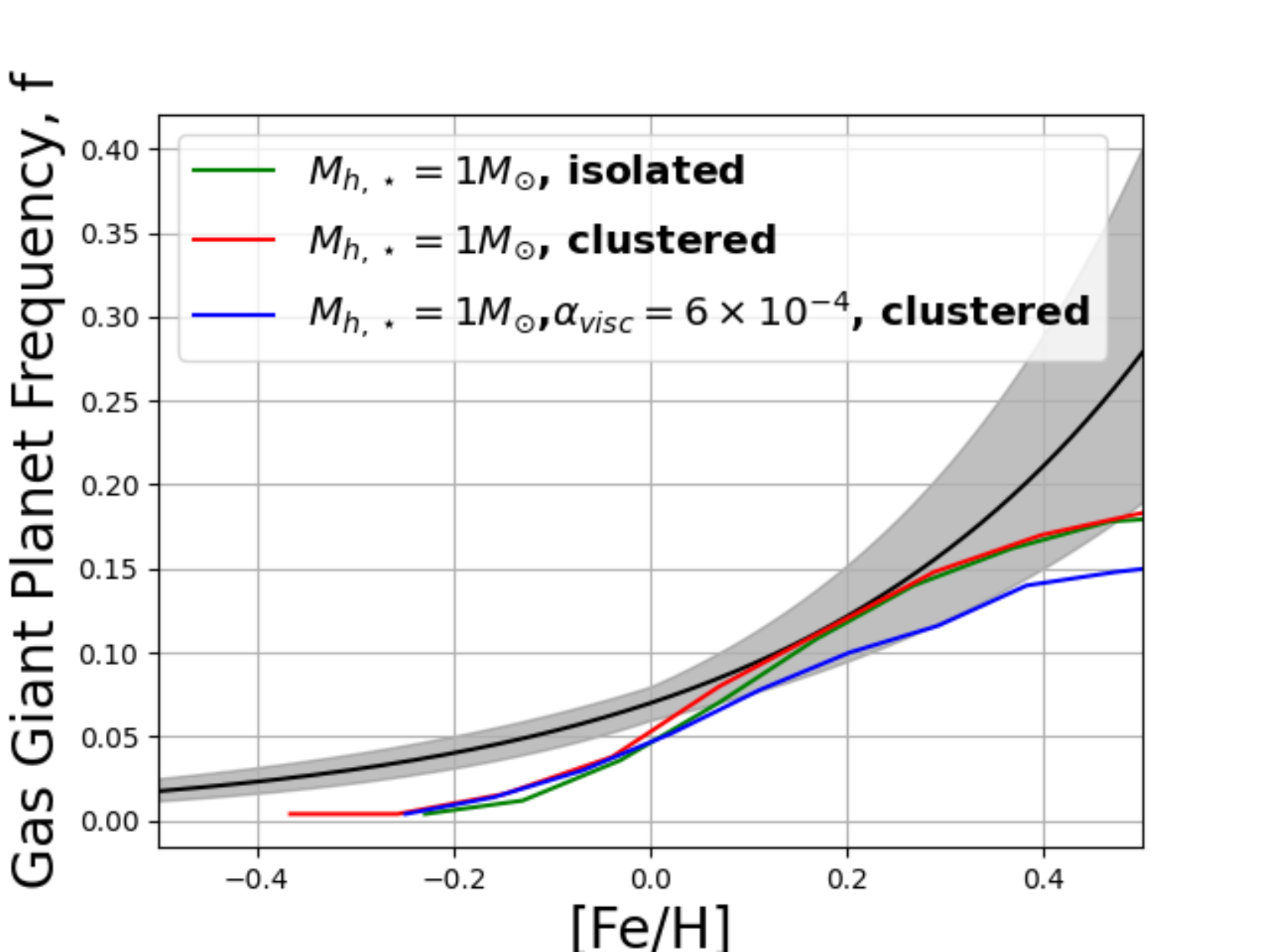}\quad
           \hfill
        \includegraphics[width=\columnwidth, 
height=2.35in]{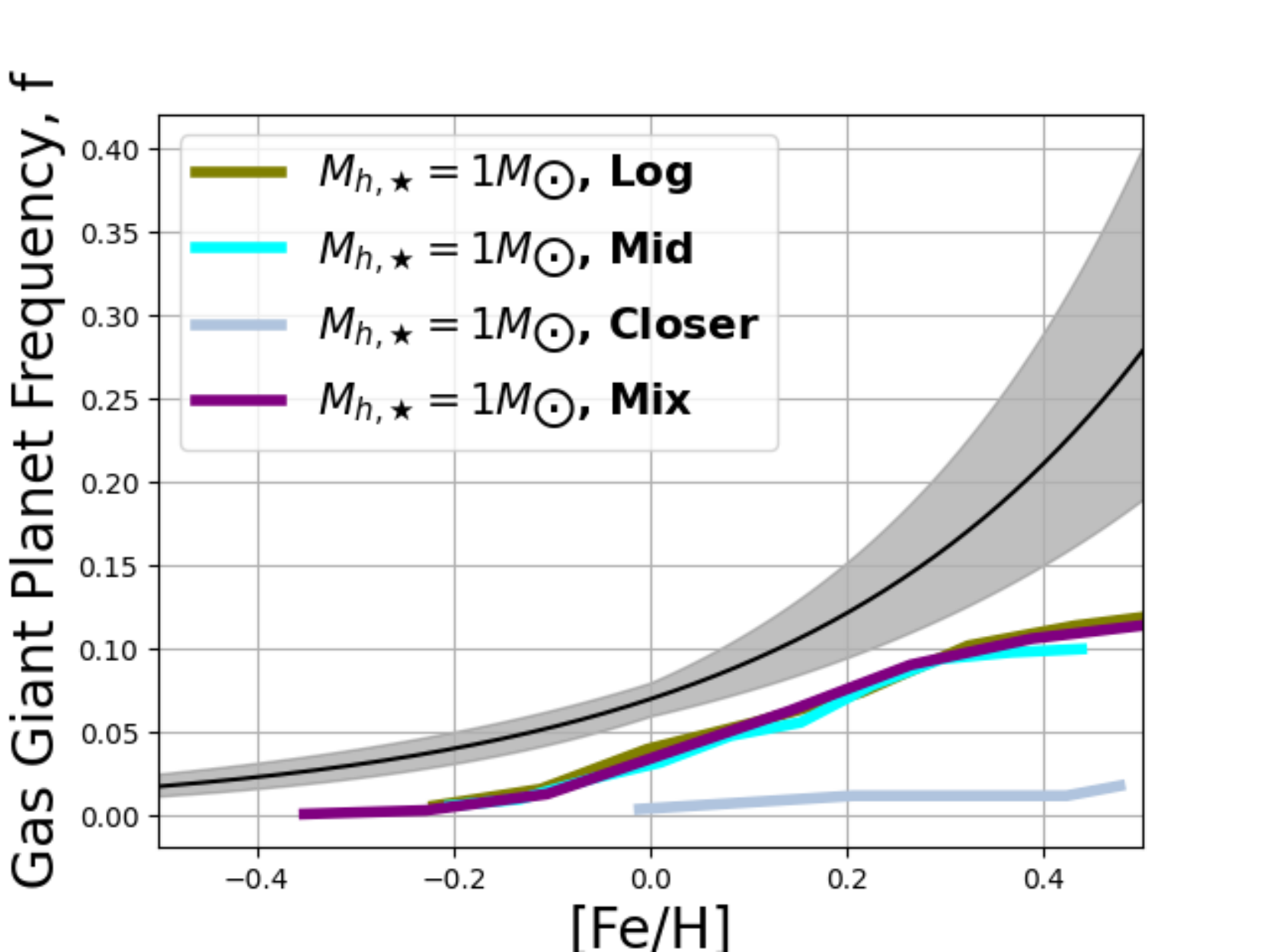}

  \caption{Giant planet frequency as function of $[F e/H]$ from the observations by \protect\cite{2010PASP..122..905J} marked by the black curve. The remaining colored curves mark the models from our simulation that represents the different stellar encounter and disc environment configurations.  The left and the right panel plots show the gas giant occurrence rates with metallicity for linear distribution and  other periastron ratio distribution (logarithmic, mid, close and mixed distributions), respectively. }
         \label{fig:6}
 \end{figure*}

\begin{figure*}
        \centering
         \includegraphics[width=\columnwidth, 
height=2.35in]{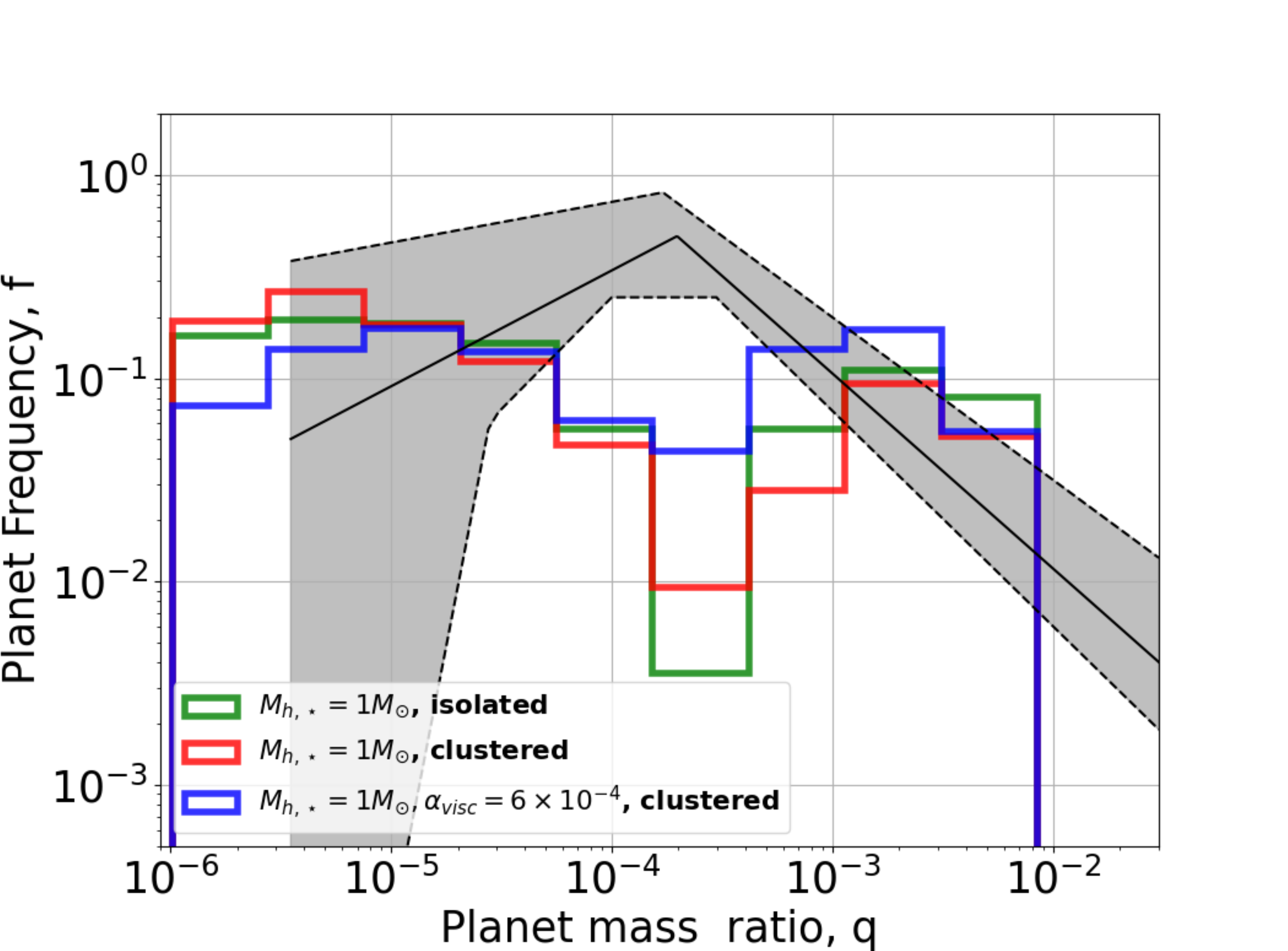}\quad
           \hfill
        \includegraphics[width=\columnwidth, 
height=2.35in]{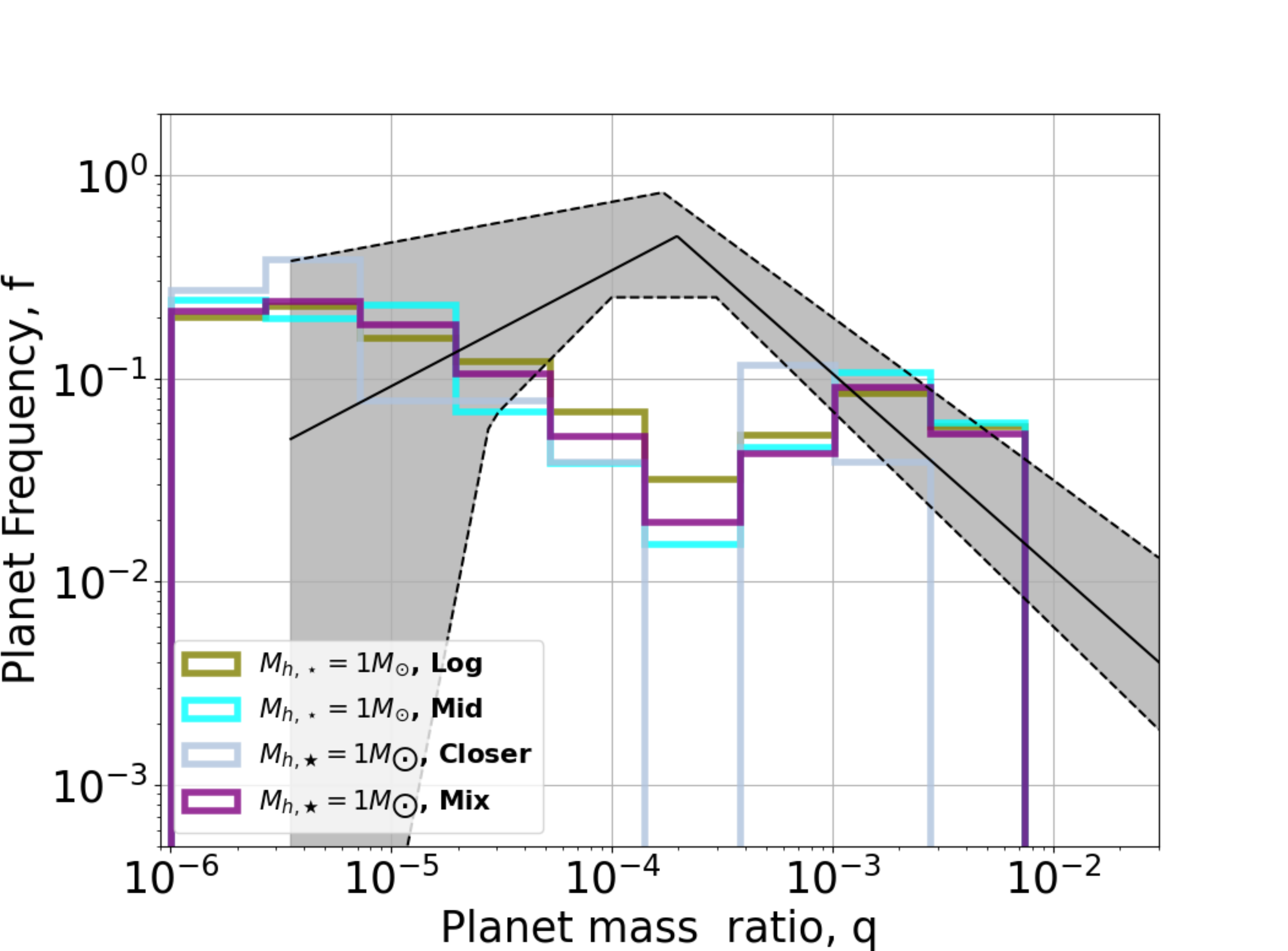}

  \caption{Comparison of our simulations to the microlensing data from 
\protect\cite{2016ApJ...833..145S}. The black solid curve depict the best planetary frequency from \protect\cite{2016ApJ...833..145S} and the grey band denote the therein  68~\% confidence interval for the best fit parameters. The comparison presented represents a subset of our synthesis data  within 2.7~au to 50~au and in the planet mass ratio, $q\gt 10^{-6}$, corresponding to the planets inside the purple square in Figure~\ref{fig:3} and Figure~\ref{fig:4}.  The left and the right panel plots show the mass occurrence rates of giant planets with linear distribution and  other periastron ratio distribution (logarithmic, mid, close and mixed distributions), respectively.}
         \label{fig:7}
 \end{figure*}

 We present in this section the comparison of the simulated gas giant planets  to a complete set of gas giant exoplanet data ($a$, $[Z/H]$ and $m$). To begin with, we show the cumulative semi-major axis distribution of our synthetic giant planet population (corresponding to grey regions in the syntheses plots) to $a$ distribution with the bias-corrected data originating from \cite{2011arXiv1109.2497M} in Figure~\ref{fig:5}. The original data from \cite{2011arXiv1109.2497M} only includes planets with $Msin(i)> 50$ Earth mass and features the orbital period instead of the semi-major axis. For our comparison, we make the assumption that these stars are solar like and convert the period into semi-major axis. Figure~\ref{fig:6} and Figure~\ref{fig:7} show the comparison of our planet formation models to the radial velocity-metallicity correlation \citep{2010PASP..122..905J} and the mass distribution from microlensing observations \citep{2016ApJ...833..145S}. The comparison is like in \cite{2018MNRAS.474..886N} and \cite{2019MNRAS.488.3625N}, where the comparison of the synthetic data to both radial velocity and microlensing field of view were performed. We note that, because the observed distributions have already been debiased, we can directly compare them to our model distributions.
 
 Overall, the models with nominal and the linear distribution in periastron ratio match the cold gas giant planets relatively well, but, clearly over predict the occurrence of hot gas giant planets, with the over prediction being less for simulations with stellar encounters (see left panel in Figure~\ref{fig:5}). In addition the over prediction reduces further for lower $\alpha$-viscosity parameter ($\alpha = 6\times10^{-4}$) due to relatively slower inward migrations of the giant planets. In absence of low $\alpha$-viscosity parameter, many planets therefore ends up assuming closer orbits, causing the over prediction of the hot gas giant planets. The reduced over predictions from the clustered simulations is because of the reduced gas giant planet formation due to the disc truncation by the stellar encounter. With our parameter configurations that gives enhanced advantage to disc truncation, the over predictions of the closed in gas giant planets are far much less. This is because the maximal disc truncation deters formation of giant planet, within the available regions of the survived discs. Similar interpretation holds for the gas giant frequency dependency on the  metallicity. There is a better match for isolated and cluster model with linear distribution in periastron ratio (see the left panel in Figure~\ref{fig:6}). However, the simulation with lower viscosity-parameter (the blue curve in Figure~\ref{fig:6}) generally reveal an underestimate of the gas giant planet occurrence due to slowed gas accretion and migration rates. Because of slowed gas accretion and migration rates, the forming planets growth rates and migration speeds are low to allow the giant planet reach the RV boundaries. The same trend is seen with stellar encounter configuration that substantially truncate the disc (the right panel plots),  where a clear under estimate of the gas giant occurrence rate is seen.
 
Peculiar for all our models is the saturation of the giant planet occurrence rate with increasing [Fe/H], see Figure~\ref{fig:6}. As [Fe/H] increases, more pebbles become available, planets accrete faster and can form more giant planets. However, at some level a saturation is reached, caused by the following two effects: (i) planetary embryos formed in the inner regions of the disc have a very low pebble isolation mass and will form super Earths that never reach runaway gas accretion, implying that an elevated metallicity will not change the fate of these planetary embryos. (ii) Planets formed in the outer regions of the disc grow faster and thus might not reach the RV detection bin, because they become gas giants very early and they migrate very slowly in type~II migration and stay outside of the radial velocity bin.

In general, the microlensing data, on the other hand, matches relatively well in the planetary mass ratio  range of $4\times 10^{-4}\leq q \leq 4\times 10^{-3}$ and for a planetary mass ratio range, $q \leq 4\times10^{-5}$ (Figure~\ref{fig:7}). However, this model does not reproduce very well the mass ratios between $4\times10^{-5}$ and $4\times10^{-4}$, consequently leaving a ``valley'' in the mass range. This behavior is very typical for planet formation simulations in the core accretion paradigm with fast gas accretion rates. Once the planet starts to accrete gas in the runaway regime it grows very fast to nearly Jupiter mass, leaving the mass range between $4\times10^{-5}$ and $4\times10^{-4}$ empty. This effect is also pointed out in \cite{Suzuki2018} where comparison of the \cite{Ida2013} models to the microlensing data was made. With moderate stellar encounter invoked in the simulation, gas giant planets in the microlensing field of view are even matched better (right panel plot in Figure~\ref{fig:7}). The $q$ dependence of the gas giant planets are better within the acceptable margins for the clustered simulations. The agreement in the distribution of the gas giant planets is because the disc truncation from clusters hinders giant planet formation, thus matching the varnishing trend of the massive gas giant planet distributions. In addition, our models do not reach high planet mass ratio end of the microlensing survey. We suspect, this problem could be because such massive planets might form via other planet formation pathways like the gravitational instability \citep{Boss1996}. Recently, \cite{Schlaufman2018} argued that planets with masses larger than 4-5 Jupiter masses are most likely formed via gravitational instabilities. However, this leaves our best model with a "gap" in the $m$- distributions of the microlensing surveys, as our simulations do not reach planetary mass ratio, $q > 10^{-2}$ (see Figure~\ref{fig:7}). However, our simulations do not include the growth and migration of multiple planets which could allow for collisions to make larger mass planets \citep[e.g.][]{Bitsch2019b}.


\section{Discussion}\label{discuss} 
 The central topic in this work is the disc truncation by stellar encounter. Various studies found that stellar flybys, in particular close stellar flybys could affect circumstellar discs and the associated planetary systems in various ways:
\begin{itemize}
 \item[(i)] Close stellar encounters could influence the structure of the planetary system \citep{2018A&A...610A..33P,2019AJ....157..125D,2020Natur.586..528W} during the early phase of planet formation history. For example, the outermost planets that form at the disc’s outskirts might experience orbital excitation due to the dynamical perturbation from the flyby. It is therefore possible, the outer planets strongly assume eccentric orbits in environment where close stellar flybys are common. Therefore the observed orbital structure of the outer planets could point at possibilities of close flybys or dynamical excitation \citep{2017A&A...598A..70S}. Orbital inclination due to orbital excitation might modify the course of planet formation; for example, pebble accretion rates and migration rates could substantially reduce \citep{2010A&A...523A..30B,2010AJ....139.1297L,2022MNRAS.510.3867C}. Circumstellar discs in stellar clusters can also be tilted after subsequent capture of gas from a dense gaseous reservoirs \citep{2011MNRAS.417.1817T}. Planets forming in these circumstellar discs might mostly assume inclined planetary orbits. Post stellar encounter discs may also have denser inner disc regions. This drives fast coagulation and consequently faster planet formation rates. It is also possible that the already forming planets gets compressed by gas inflows leading to a hot Jupiters and short-period eccentric planets \citep{2011MNRAS.417.1817T}. Self-consistently integrating this important aspects of planet formation to global planet formation frameworks, i.e the planet population syntheses is computationally demanding. For this reason we did not model orbital excitation of planets due to stellar flyby. With orbital excitation due to stellar encounter implemented in our model, we envision a much more reduced gas giant planet formation rates, in particular the cold gas giant planets formation rates. Nevertheless, the syntheses calculations in this study point out clearly and in realistic manner the global contribution of disc truncation in planet formation.
\item[(ii)] Close stellar flybys could result into formation of substructures (rings) in discs. Because rings in discs are hotspot for enhanced dust densities—boost for dust coagulation and growth-consequently planet formation \citep{2020A&A...638A...1M}, flybys could promote planet formation. Global dust evolution and planet formation  in such ring structures have been studied in detail in our submitted paper \citep{Andama_submitted_II}. Formation of planetary cores in such ring structure allows for a modification in core accretion paradigm, for example, the solid core mass may not be limited by the famous isolation mass but to the total amount of dust trapped in the ring structure \citep{2020A&A...638A...1M,Andama_submitted_II}. Via this phenomenon, we believe that stellar encounter could be a blessing in disguise in enhancing planet formation in stellar clusters. However, it remains questionable, how often the stellar encounter lead to formation of substructures in the host disc. Formulation of this phenomenon is still under development \citep{2021ApJ...921...90P} and therefore difficult at this stage to self-consistently anchor to global planet formation models.  Although calculating substructures by stellar flyby is clearly beyond the scope of this work, it is interesting in future to; calculate in detail the frequency of such substructures due to stellar flybys, estimate the collective impact of disc truncation and substructures formation by stellar flybys on planet formation.

\item[(iii)] Protoplanetary discs can in principle be truncated by an encounter with a visiting star in a dense stellar environment. \cite{1993MNRAS.261..190C} showed that disc could be truncated up to one-third of the distance of the encounter. For example, if the closest approach of a star were 300 AU, then the disc would be truncated to a radius of 100 AU \citep{1996MNRAS.278..303H}. Further simulations showed that prograde, co-planar encounters are the most destructive to the disc \citep{1998MNRAS.300.1189B}. It is also possible material from the disc can also be transferred onto an orbit around the perturbing star \citep{2001Icar..153..416K}. In this study, effect of disc truncation is calculated by following the disc truncation fitting presented in \cite{2014A&A...565A.130B}. As in \cite{2014A&A...565A.130B}, we assumed; the most destructive encounters; parameter configurations that mimics the ONC parameter space and simplistic case of the disc harbouring only the host star, but not the perturbing stars\footnote{Such assumption are also seen in \cite{2005A&A...437..967P} that aims to reduce computational burdens of such complex physical processes.}. Anchoring the earlier disc truncation results \citep{1993MNRAS.261..190C,1996MNRAS.278..303H} into our global planet formation model would report less disc truncation effect than reported here. For example, the gas giant planet population might be enhanced compared to this study.

\item[(iv)] Close flybys can lead to gravitational instabilities that might eventually trigger planet formation \citep{2010ApJ...717..577T,2020RSOS....701271P}. Although the most acceptable pathway of planet formation is thought to be the core accretion scenario \citep[e.g,][]{wetherill1980,kokubo1998,thommes2003,coleman2014},  planets could also form from the fragmentation of thermodynamically cold circumstellar discs \citep{kuiper1951, cameron1978,boss1997,gammie2001,rice2003,tanga2004,rafikov2005,durisen2006,2002Sci...298.1756M}. Suppose planet form via fragmentation, then interactions with passing stars during the protoplanetary disc stage could suppress planet formation \citep{2009MNRAS.400.2022F}. Furthermore, some studies find that in discs affected by stellar flybys, the planets that form are more massive and at larger orbital distances \citep{2009A&A...505..873F}. Because our focus is in the explanation of the broad gas giant planet sub population, we did not model planet formation via gravitational instability paradigm that favours the formation of cold gas giant planets, but instead follow core accretion formalism that proved to reproduce broad range of planetary systems quite well. 
\end{itemize}
This work assumed a stringent condition of one encounter with passing stars per planet formation history. In principle, stars can undergo multiple stellar interactions that might drastically destroy the host discs \citep{2020RSOS....701271P}. Encounter rate might vary with cluster evolution, for example study by \cite{2018A&A...610A..33P} found close flybys should be common after cluster formation in Pr 0211 systems with a close fly-by rate of around 0.2-0.5~$\rm My^{-1}$. With this fly-by rates, it means that for disc with lifetime of 2-3~My (most frequent disc lifetime in our model), roughly 1 encounter would be registered. This is in agreement with our estimations of 1 encounter per planet formation history. For higher disc lifetimes, it is therefore possible for at least two encounters per planet formation history. However, such long disc lifetimes are rare in our simulations. It is therefore a realistic approximation to assume one encounter per planet formation history. We have however acknowledge that through multiple encounters, the suppression of gas giant planet formation is enhanced.

\section{Conclusion}\label{Conclusion} 

In this work, we explore the outcome of our pebble-based core accretion planet formation models with pebble sizes corresponding to two population dust distribution of \cite{2012A&A...539A.148B} and with the setup of disc truncation from stellar encounters following \cite{2014A&A...565A.130B}. We analysed the emergent proportion of the gas giant sub populations and compared the synthesized population; with the  debiased $a$ distribution extracted from the HARPS survey \citep{2011arXiv1109.2497M}, the planetary mass distribution determined by microlensing \citep{2016ApJ...833..145S} as well as with the giant planet occurrence rate (frequency-metallicity relation) determined by RV observations \citep{2010PASP..122..905J}. We focused on the giant planet population, not the super earth planets because their formation requires proper inclusion of the N-body paradigm which is beyond the scope of this work.

Compared to our previous model \citep{2018MNRAS.474..886N}, we have included a different approach on gas accretion \citep{Crida2017,2017EPSC...11...44C,2020A&A...643A.133B,2021MNRAS.501.2017N} and a different approach for type~I \citep{2014MNRAS.444.2031P} as well. Besides the stellar encounter parameter sampling, the randomization of the important parameters in our model (e.g. disc metallicity and starting times) were similar as in our previous work. Our planet formation simulations are limited to only one planetary embryo and only one stellar encounter per simulation. 

Using the debiased $Msin(i)-P$ data from \cite{2011arXiv1109.2497M}, we identified that our models result in too fast inward migration of the giant planets (Figure~\ref{fig:5}). The fast inward migration due to the large viscosity also conform to the idea that Jupiter's core could have formed beyond 20~au \citep{2015A&A...582A.112B,Oberg2020}. Nevertheless, our planet formation simulations indicate that relatively more cold Jupiters than hot Jupiter form in discs without stellar encounter compared to discs that experienced stellar encounters via linear periastron ratio distribution. Our simulations also show that the ratio of hot to cold Jupiter depends on the periastron ratio distribution (see Table~\ref{tab:1}) that somehow determines the severity of the disc truncation in our stellar encounter calculation (Figure~\ref{fig:2paramImpact}).

In addition, we also compared our simulation data to the giant planet occurrence with metallicity of RV data (Figure~\ref{fig:6}) and the mass distribution curve of the microlensing data (Figure~\ref{fig:7}). The isolated disc simulations reproduces the metallicity-giant planet correlation quite well, but the models though predict a shallower metallicity dependence of giant planets than reflected by observations for simulations with disc truncation. In addition, the models also over predicts the $a$ dependence of the giant planet semi-major axis distribution for smaller semi-major axis ($a \leq 3$~au). The overestimate of the $a$ dependence of giant planet could be reduced via inviscid disc environments and stellar cluster environments. We reported relatively acceptable comparisons of the simulation results to the microlensing data for planet mass ratio range, $4\times 10^{-4}\leq q \leq 4\times 10^{-3}$ and for a planetary mass ratio range, $q \leq 4\times10^{-5}$ (Figure~\ref{fig:7}). The stellar cluster encounter incorporation solved the over fitting challenge for the massive gas giant planets.

 Depending on the strength of the disc truncation, stellar encounters in general lowers in the simulations the occurrence rates of:
\begin{itemize}
 \item [--] cold Jupiters due to the truncation of the disc size that shifts the planet formation channels closer to the host star.
 \item [--] hot Jupiters because of reduction in the host disc mass from the effect of stellar encounter.
\end{itemize} Since the hot Jupiters and the cold Jupiters in our models preferentially form at a closer and at a farther location from the host stars. It is possible that in our simulations the disc truncation on the host disc's characteristic radius is an ambient one. Otherwise, an immense disc truncation on the host disc would mostly leave formation regions that favor hot Jupiters formation, not the cold Jupiters formations. Previous study also found out that not many of the host stars in clusters are significantly impacted by stellar clusters. For example, \cite{2018A&A...610A..33P} revealed that stellar encounter might affect only 12-20~\% of protoplanetary discs.

Due to computational cost, it was not possible at the moment to study the influence of evolution of some of the disc parameters, i.e the $\alpha$-viscosity parameter, which we envision to have contribution to the extent of disc truncation through disc spreading and also impact on the shape of the grain size distributions. Of utmost importance are; the background heating and photo-evaporation effect of stellar clusters, other impacts of stellar encounter-particularly the substructures formations. Though very complex, we expect that the simultaneous occurrence of these cluster phenomena in planet formation simulations for the clusters' lifetimes present even more reasonable stand. Similarly, the opacities in the planetary envelopes in our simulations are not self-consistently calculated from the disc but rather fixed ($\kappa_{\rm env}=0.05~\rm cm^{2}g^{-1}$) . Envelope opacity impacts envelope cooling rates and thus gas accretion rates \citep{Mordasini2014}. We also see substantial potential for improvement in the treatment of the core chemical compositions that could improve the overall core accretion rates \citep{2021arXiv210513267S,2021arXiv210903589S}.

Within the limitation of this model, we conclude that inclusions of stellar encounters calculation influences structures of protoplanetary discs, hence planet formations and the outcome of global models, like planet population synthesis models. To reinforce our conclusion, more detailed follow-up planet population syntheses simulations are needed to concurrently test the impact of simultaneous inclusion of the nexus of the physical phenomenons in stellar clusters. More detailed simulations are also needed to account for the impact of competitions for the multiple cores on the work presented herein.

\subsection*{Acknowledgments}
 
 We acknowledge funding from the Center for Space Research (CSR)–North West University, South Africa. N.N, thanks Bertram Bitsch for the fruitful discussion at the early stage of the manuscript. We thank an anonymous referee whose comments helped to improve this manuscript.

\subsection*{Data Availability} 
 The codes used to obtain the results in this paper is available upon reasonable request.

\bibliography{./biblio.bib}
\bibliographystyle{mnras}
\bsp 
 
\label{lastpage}
\end{document}